\documentclass[letterpaper,11pt]{article}
\usepackage{jheppub}
\usepackage{comment}
\pdfoutput=1
\usepackage{dcolumn}
\usepackage{bm}
\usepackage{graphicx}
\usepackage{amssymb,amsmath}
\usepackage{multirow}
\usepackage{color,url}
\usepackage{tabu}
\usepackage{array}
\usepackage{ulem}
\usepackage{soul}
\usepackage{colordvi}
\usepackage{subcaption}
\usepackage{ mathrsfs }
\def\ga{\mathrel{\raise.3ex\hbox{$>$\kern-.75em\lower1ex\hbox{$\sim$}}}}
\def\la{\mathrel{\raise.3ex\hbox{$<$\kern-.75em\lower1ex\hbox{$\sim$}}}}
\def\beqa{\begin{eqnarray}}
\def\eeqa{\end{eqnarray}}

\begin{document}

\title{\boldmath Disentangling Boosted Higgs Boson \\ Production Modes with Machine Learning }

\author[1]{Yi-Lun Chung,}
\author[2]{Shih-Chieh Hsu,}
\author[3]{and Benjamin Nachman}
\affiliation[1]{\normalsize Department of Physics, National Tsing Hua University, Hsinchu 300, Taiwan}
\affiliation[2]{\normalsize Department of Physics, University of Washington, Seattle, Washington 98195, USA}
\affiliation[3]{\normalsize Physics Division, Lawrence Berkeley National Laboratory, Berkeley, CA 94720, USA}

\emailAdd{s107022801@m107.nthu.edu.tw}
\emailAdd{schsu@uw.edu}
\emailAdd{bpnachman@lbl.gov}

\date{\today}

\abstract{
Higgs Bosons produced via gluon-gluon fusion with large transverse momentum ($p_T$) are sensitive probes of physics beyond the Standard Model. However, high $p_T$ Higgs Boson production is contaminated by many production modes other than gluon-gluon fusion, including vector boson fusion, production of a Higgs boson in association with a vector boson, and production of a Higgs boson with a top-quark pair.  By using modern machine learning techniques to analyze jet substructure and event information, we demonstrate the capability of machine learning to identify production modes accurately. These tools hold great discovery potential for boosted Higgs bosons produced via ggF, and may also provide additional information about the Higgs Boson sector of the Standard Model in extreme regions of phase space for other production modes.
}

\maketitle

\newpage
\section{Introduction}\label{introduction}
Since the Higgs boson was discovered at the Large Hadron Collider (LHC) in 2012, it has been used as a tool in the search for physics beyond the Standard Model (BSM).  One particularly sensitive channel is the loop-induced gluon-gluon fusion (ggF) mode $gg\to\mathrm{H}$ at high transverse momentum ($p_T$).  Due to the large Higgs boson branching ratio to bottom quarks, $\mathrm{H\to b\bar{b}}$, there is a particularly interesting topology for studying high-$p_T$ Higgs bosons \citep{ATLAS:2018hzj,Sirunyan:2020hwz}. Multiple groups have developed dedicated techniques for distinguishing $\mathrm{H\to b\bar{b}}$ from the large multijet background dominated by $g\to b\bar{b}$~\cite{Aad:2019uoz,CMS-DP-2018-046,Datta:2017lxt,Lin:2018cin,Moreno:2019neq}. However, the high $p_T$ Higgs boson cross section has significant contributions from processes other than ggH, including vector boson fusion (VBF), vector-boson associated production (VH), and top-quark pair associated production (ttH) \cite{Becker:2020rjp, Pagani:2020rsg}. If $gg\to \mathrm{H}$ could be clearly separated from the other Higgs production modes, the sensitivity to BSM would be enhanced.

State-of-the-art machine learning (ML) techniques have great potential to enhance the physics program at the LHC by effectively using low-level, high-dimensional information~\cite{Larkoski:2017jix,Guest:2018yhq,Albertsson:2018maf,Radovic:2018dip,Carleo:2019ptp,Bourilkov:2019yoi}. One such technique is to treat collider events as images and reconstruct events from detector data using a deep convolutionql neural network (CNN)~\cite{Pumplin:1991kc,Cogan:2014oua,Almeida:2015jua,deOliveira:2015xxd,ATL-PHYS-PUB-2017-017,Lin:2018cin,Komiske:2018oaa,Barnard:2016qma,Komiske:2016rsd,Kasieczka:2017nvn,Macaluso:2018tck,li2020reconstructing,li2020attention,Nguyen:2018ugw,ATL-PHYS-PUB-2019-028,Andrews:2018nwy,Ngairangbam:2020ksz}.  One can combine event-level and region-of-interest information into a two-stream CNN (2CNN) that has been shown to effectively separate boosted $H\rightarrow b\bar{b}$ from $g\rightarrow b\bar{b}$ in simulation~\cite{Lin:2018cin}.  This architecture is the inspiration for the present work in which a similar method is deployed to investigate the separation of the various Higgs boson production modes at high $p_T$.  For the purposes of this study, a boosted decision tree (BDT) will serve as a baseline classifier for comparison with the 2CNN.

This paper is organized as follows. The simulated samples are described in Section \ref{mc_samples}. Section \ref{machine_learning_classifiers} describes the ML setup, including the 2CNN architecture and high-level features for the BDT and image preprocessing. The results are presented in Section \ref{results} with a discussion about the information used by the 2CNN in Section \ref{discussion}. Conclusions are presented in Section \ref{conclusions}.

\section{Monte Carlo Samples} \label{mc_samples}
This study considers the four main Higgs production mechanisms: ggF, VBF, VH, and ttH. The program {\texttt{M{\footnotesize{AD}}G{\footnotesize{RAPH}}5\_{\footnotesize{A}}MC@NLO 2.7.2}}~\citep{Alwall:2014hca} is used for modeling $pp$ collisions at $\sqrt{s}$ = 13 TeV and at next-to-leading order (NLO) accuracy in the strong coupling. The {\texttt{PDF4LHC15\_nnlo\_mc}} \citep{Botje:2011sn} parton distribution function (PDF) is used. The hard-scattering events are passed to {\texttt{P{\footnotesize{YTHIA}} 8.244}}~\citep{Sjostrand:2007gs} to simulate the parton shower and hadronization, using the default settings and {\texttt{FxFx}}~\citep{Frederix:2012ps} matching applied with a merging scale of 30 GeV. In all four production modes, the Higgs boson is set to decay into b$\bar{\mathrm{b}}$ 100\% of the time.  The vector boson and top quark decay hadronically for the VH and ttH production modes, respectively.

The ggF production mode is generated with up to two jets in the matrix element using the Higgs Characterisation model~\citep{Artoisenet:2013puc,Demartin:2014fia}. For including finite top mass effect, the ggF samples are normalized to the cumulative cross section given in Ref.~\citep{Becker:2020rjp}. The additional processes are generated with up to one additional jet in the matrix element.  When both QCD and EW corrections are considered, the samples are normalized to the cumulative cross section given in Ref.~\citep{Becker:2020rjp}.

{\texttt{F{\footnotesize{AST}}J{\footnotesize{ET}} 3.2.1}} \citep{Cacciari:2011ma} and the {\texttt{F{\footnotesize{AST}}J{\footnotesize{ET}} {\footnotesize{CONTRIB}}}} extensions are used to cluster events. Following the procedure in Ref.~\citep{Lin:2018cin}, jets are clustered with the $R = 0.8$ anti-$k_t$ algorithm~\citep{Cacciari:2008gp}. These large-$R$ jets are groomed using the soft-drop algorithm~\citep{Larkoski:2014wba,Dasgupta:2013ihk} with $\beta$ = 0 and $z_{cut}$ = 0.1. The Higgs jet is required to satisfy $p_T$ $>$ 400 GeV, $|\eta|$ $<$ 2, invariant mass $>$ 50 GeV and to satisfy double $b$-tagging. Jets are declared double $b$-tagged if they have two or more ghost-associated~\citep{Cacciari:2008gn,Buckley:2015gua} $B$ hadrons with $p_T$ $>$ 5 GeV. Moreover, the leading double-$b$-tagged jet is required to have -6.0 $<$ $\rho$ $<$ -2.1 ($\rho=\mathrm{log}(m_\mathrm{SD}^2/p_T^2)$, where $m_\mathrm{SD}$ is the soft-drop mass) to avoid the nonperturbative regime of the soft-drop mass distribution and instabilities due to finite cone limitations from jet clustering~\citep{Sirunyan:2017dgc}. Finally, to be consistent with Ref.~\citep{Lin:2018cin}, a cut on the two-prong observable $N_2$~\citep{Moult:2016cvt} $<$ 0.4 is applied. This observable provides excellent discrimination between two-prong signal jets and QCD background jets.

\begin{figure}[h]
\centering
     \begin{subfigure}{0.45\textwidth}
        \centering
        \includegraphics[width=2.5in]{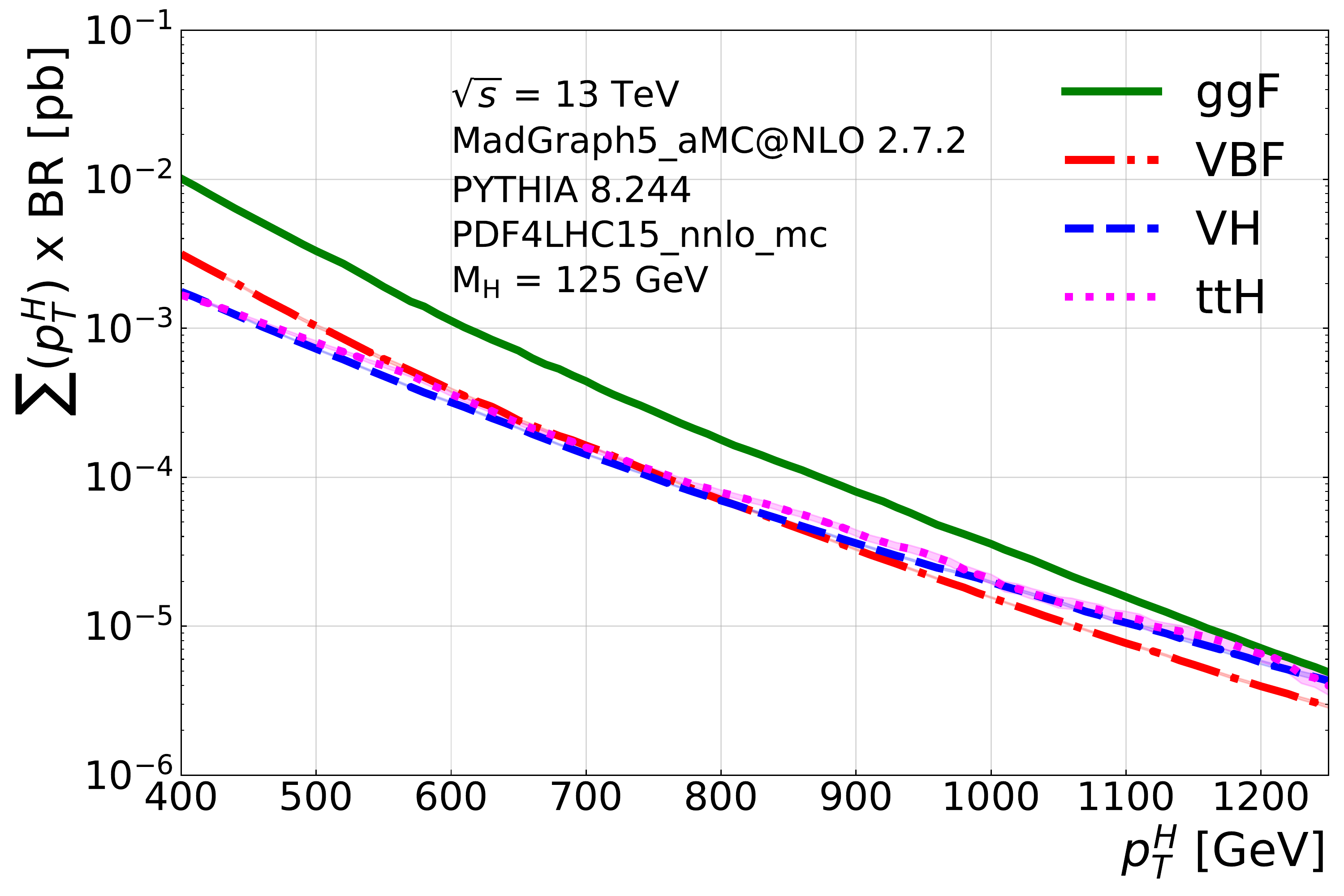}
     \end{subfigure}
     \begin{subfigure}{0.45\textwidth}
        \centering
        \includegraphics[width=2.5in]{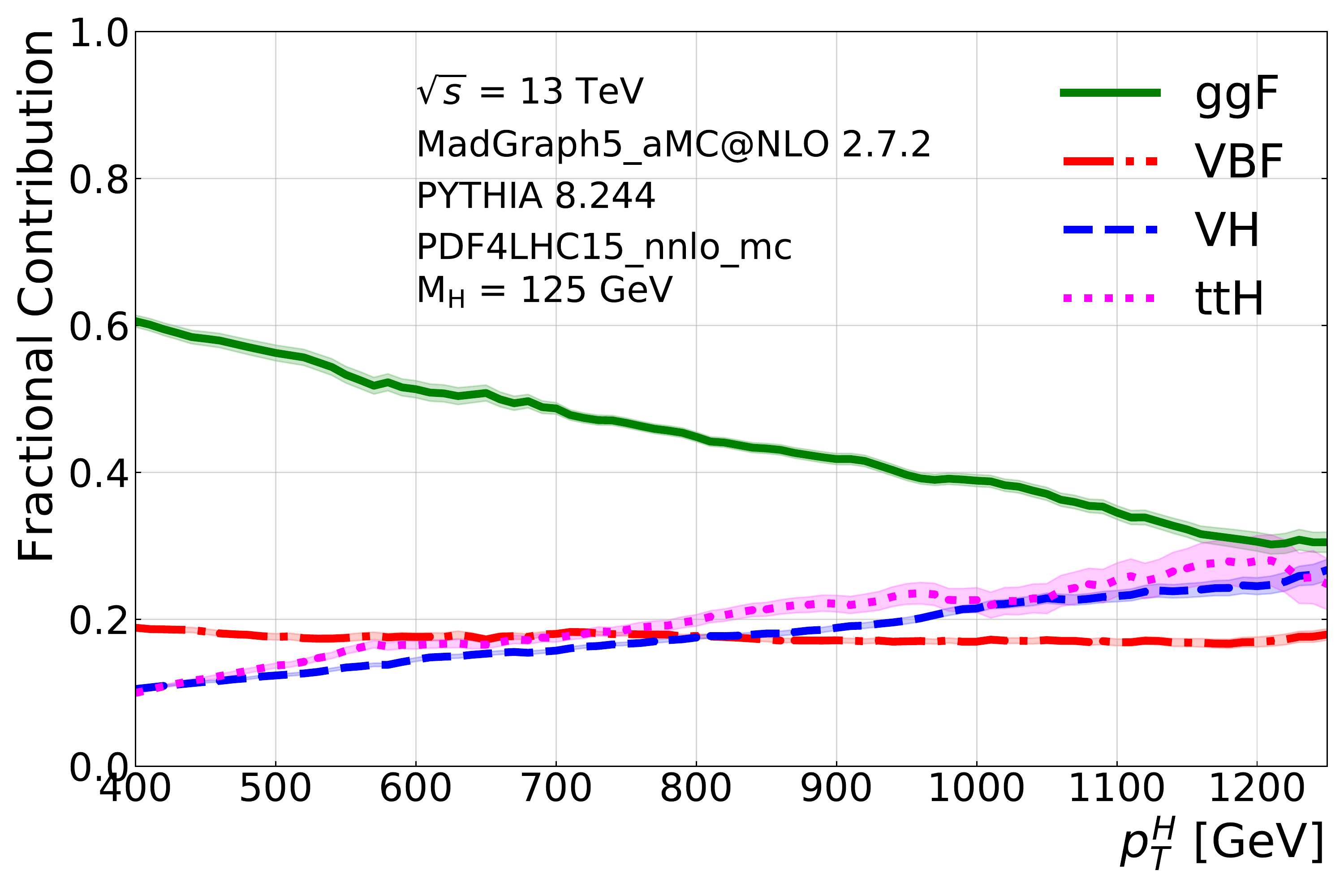}
     \end{subfigure}
\caption{The cumulative cross section (left) and fractional contribution (right) of the four Higgs production mechanisms after preselection.  The lighter error band of each production mode represents the one $\sigma$ statistical uncertainty.}
\label{fig:CumulativeXection}
\end{figure}

Figure~\ref{fig:CumulativeXection} shows the distribution of cumulative cross sections and fractional ggF contributions after applying the preselection described above. These two plots include the branching ratio $\mathcal{B}$(H$ \to$ b$\bar{\mathrm{b}}$)= 58.24\% with $\mathrm{M_H} = 125$~GeV~\citep{deFlorian:2227475},  $\mathcal{B}$(W$ \to$ hadrons) = 67.41\% \citep{Tanabashi:2018oca} and  $\mathcal{B}$(Z$ \to$ hadrons) = 69.91\% \citep{Tanabashi:2018oca}. The cumulative cross section~\citep{Becker:2020rjp} is defined as

\begin{equation}
    \sum (p^{cut}_T)=\int^{\infty}_{p^{cut}_T}\frac{d\sigma}{dp'_T}dp'_T\,.
\end{equation}

To generate enough statistics for training and testing in whole $p_T$ range, the samples are generated in three $p_T$ slices, namely 350 GeV $<$ $p_T$ $<$ 700 GeV, 700 GeV $<$ $p_T$ $<$ 1000 GeV and 1000 GeV $<$ $p_T$. After applying the preselection, there are 170k, 25k and 229k events from each production mode for training, validation and testing, respectively. For each production mode, there are 30\% in the first $p_T$ slice, 40\% in the second $p_T$ slice and 30\% in the last slice.

The sizable testing events can decrease the statistic uncertainty in the cumulative cross section and fractional contribution.

\section{Machine Learning Classifiers}\label{machine_learning_classifiers}

In this section, two ML classifiers are used to disentangle the four Higgs production modes. The first classifier is a BDT. This algorithm combines the advantages of gradient descent and decision trees~\citep{FRIEDMAN2002367,Ridgeway2006GeneralizedBM,Chen_2016,NIPS2017_6907}.
The second classifier is the 2CNN ~\citep{Lin:2018cin}. This network uses the full-event information and local information about the leading non-Higgs jet to separate the four Higgs production modes. The leading non-Higgs jet plays an important role in both the high-level and low-level information. This jet will simply be referred to as `the leading jet' in the following.

The BDT and 2CNN are trained as four-dimensional functions with one-hot encoding: (1,0,0,0) for ggF, (0,1,0,0) for VBF, (0,0,1,0) for VH, and (0,0,0,1) for ttH. In these four-component vectors, each component will represent the probability of a given mode given the input features.  These probabilities will be denoted p(mode), where mode is one of ggF, VBF, VH or ttH.

\subsection{The Boosted Decision Tree}

In this study, the BDT uses Gradient Tree Boosting. It has a fixed number of estimators (500)  with maximum depth 2. The minimum number of samples is fixed at 25\% as required to split an internal node and 5\% as required to be at a leaf node. The deviance of loss function with the learning rate 0.3. This BDT model is trained on the high-level features of the jet using the {\texttt{scikit-learn}} library \cite{scikit-learn}. 

In order to distinguish the four Higgs production modes via the BDT, the following five high-level features \citep{Belyaev_2017,Chen:2017abe,Shelton:2013an} are used in the BDT analysis: \\

\quad
1. $M_J$ : the invariant mass of the leading jet;

\quad
2. $\eta_J$ : the pseudorapidity of the leading jet;

\quad
3. $|\Delta\eta_{JJ}|$ : the absolute $\eta$ difference between the leading and subleading jets;

\quad
4. $M_{JJ}$ : the invariant mass of two leading jets ;


\quad
5. girth $\, \equiv \sum^N_{i\in J}\frac{p^i_{T}r_i}{p^J_T}$ : the girth summed over the leading jet;

\quad
6. $\Psi \equiv \frac{1}{N}\sum^N_{i\in J} \frac{p^i_{T}(0<r_i<0.1)}{p^J_T}$: the central integrated leading-jet shape,\\

\noindent where the $J$ is the leading jet, the $p^i_{T}$ is $p_T$ of the $i^\text{th}$ constituent in $J$, $r_i$ is the radius distance between $i^\text{th}$ constituent and the $J$ axis, and $N$ is the number of constituents in $J$. \\

\begin{figure}[ht!]
\centering
    \begin{subfigure}{0.43\textwidth}
    \centering
    \includegraphics[width=1.\columnwidth]{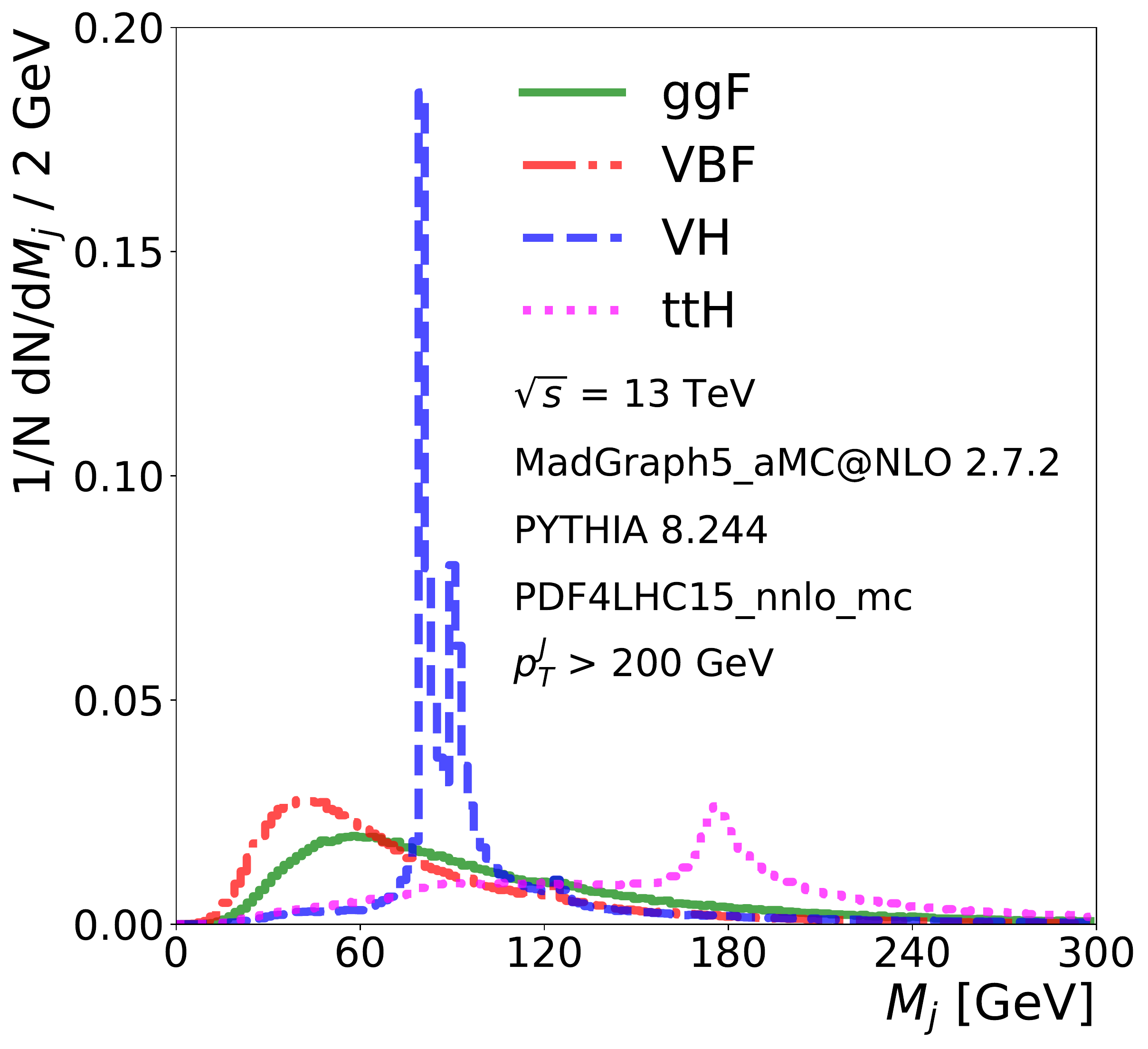}  
    \end{subfigure}
    \begin{subfigure}{0.43\textwidth}
    \centering
    \includegraphics[width=1.\columnwidth]{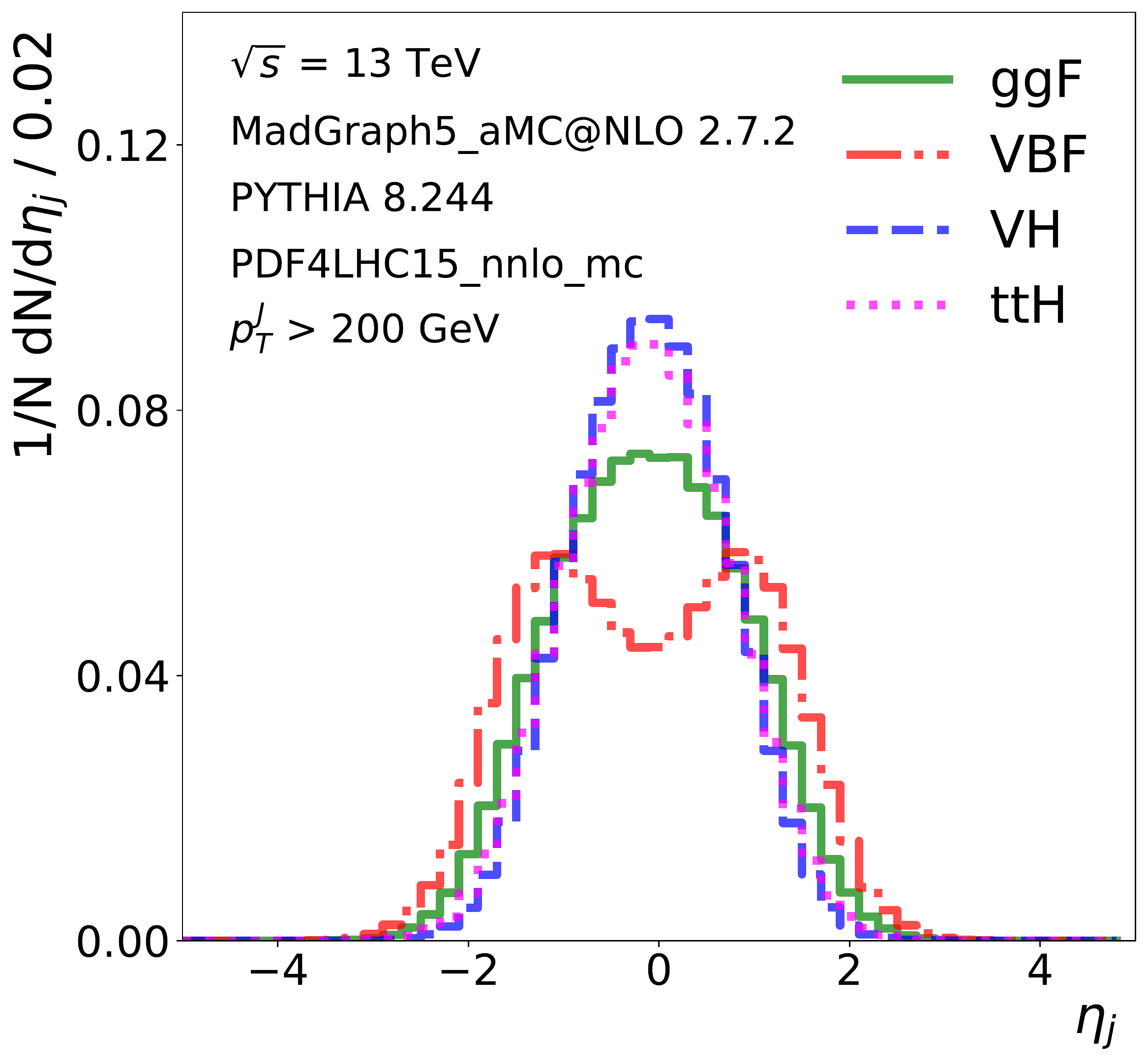}
    \end{subfigure}
    \begin{subfigure}{0.43\textwidth}
    \centering
    \includegraphics[width=1.\columnwidth]{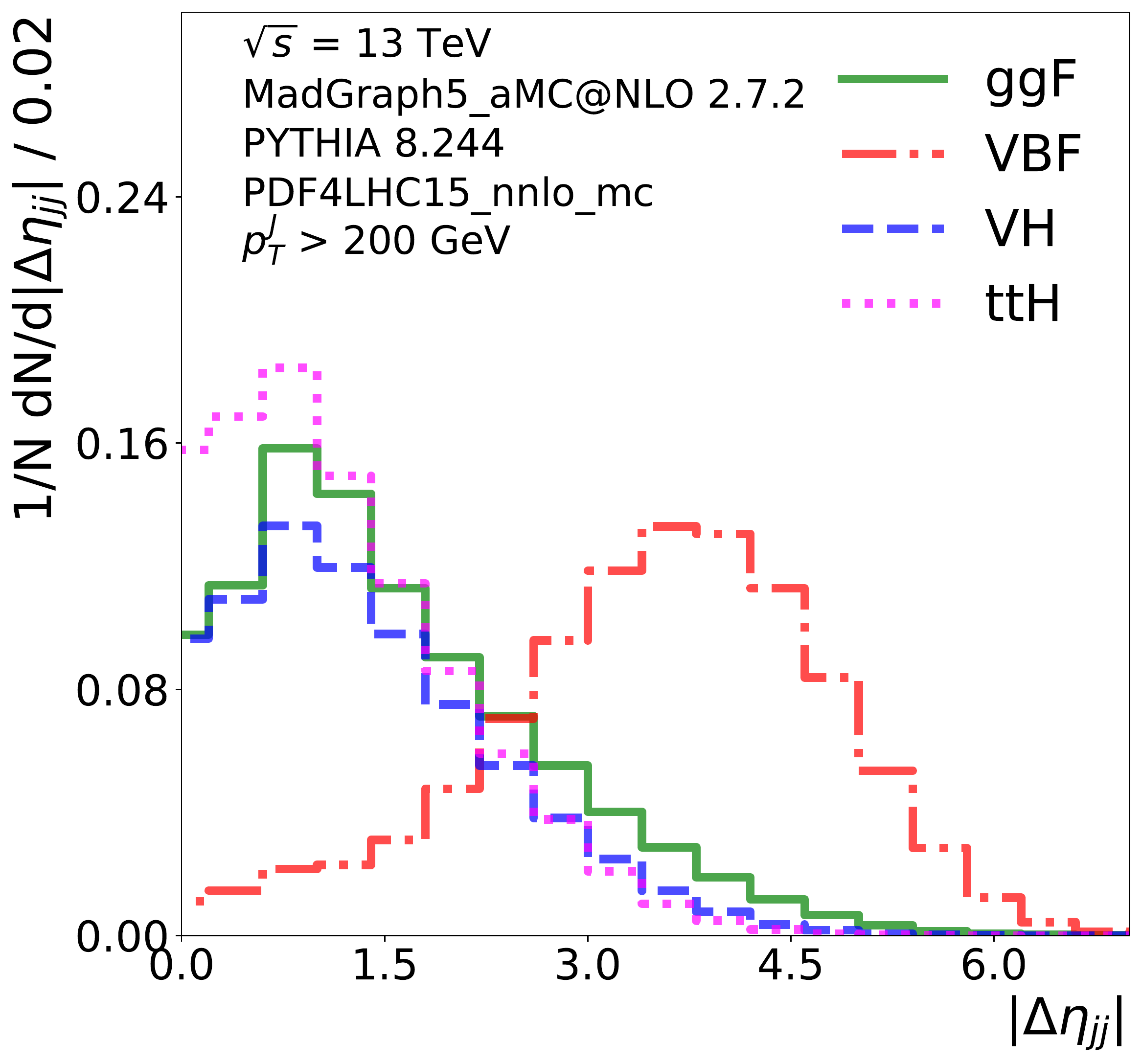}
    \end{subfigure}
    \begin{subfigure}{0.43\textwidth}
    \centering
    \includegraphics[width=1.\columnwidth]{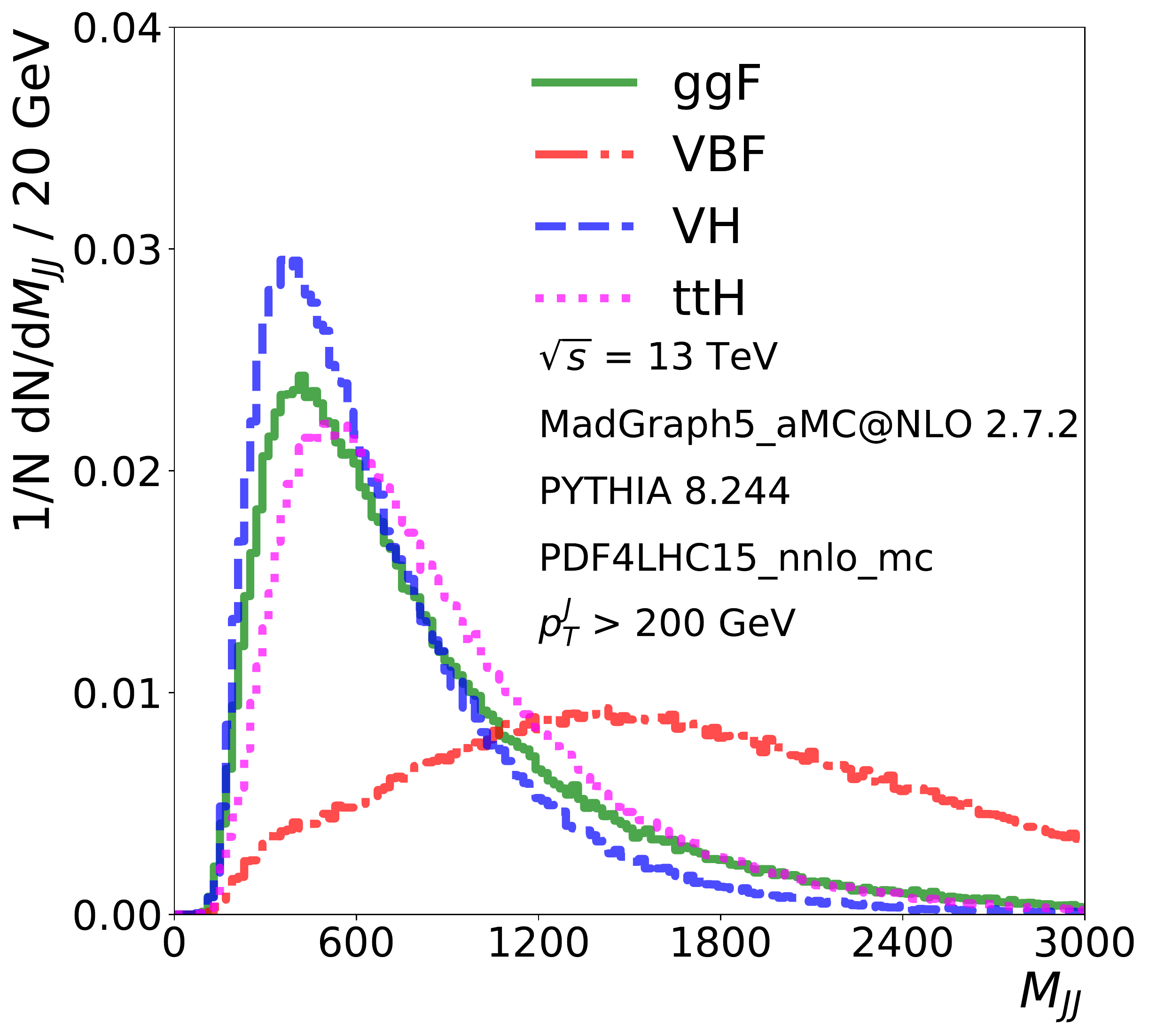}
    \end{subfigure}
    \begin{subfigure}{0.43\textwidth}
    \centering
    \includegraphics[width=1.\columnwidth]{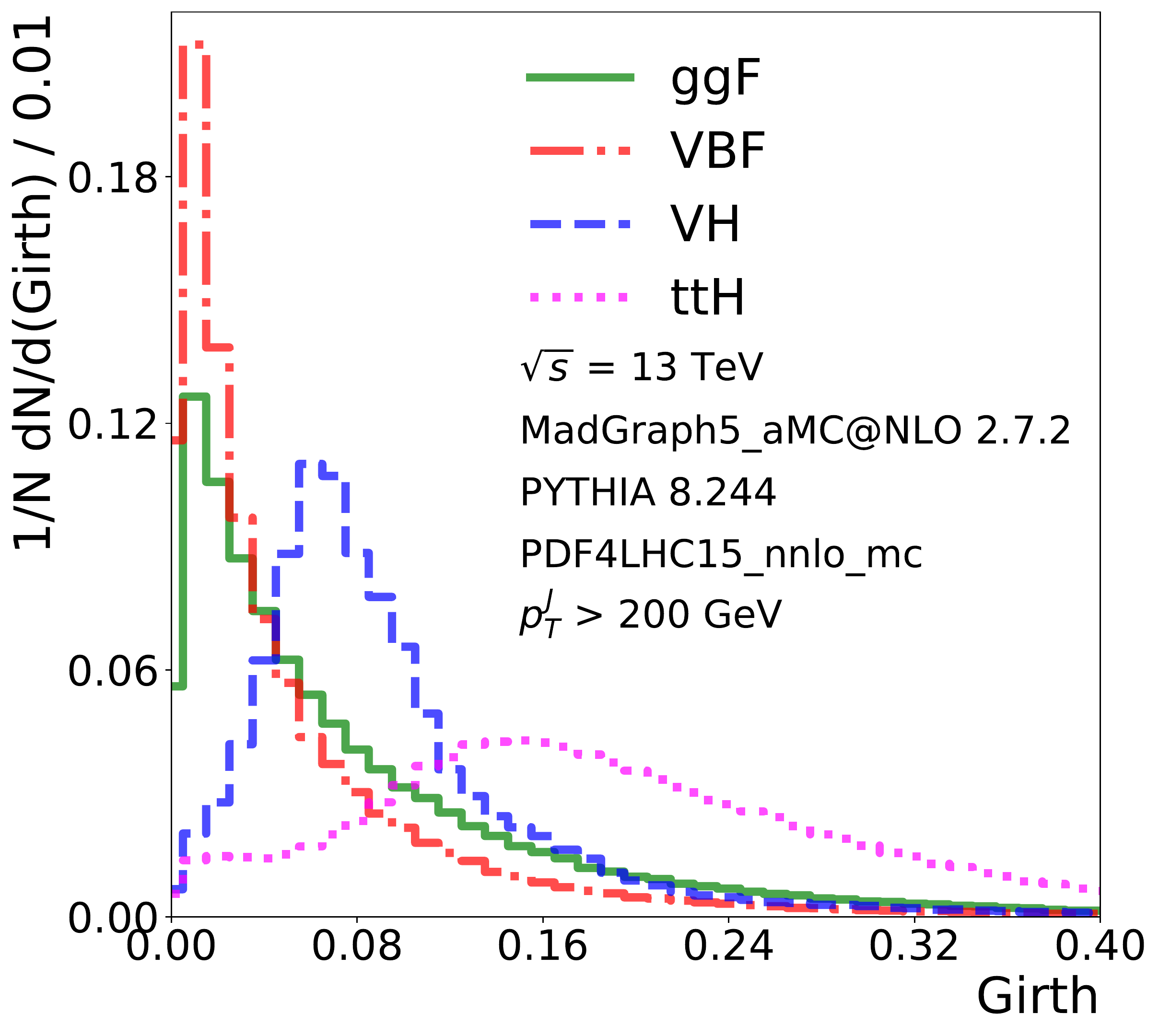}
    \end{subfigure}
    \begin{subfigure}{0.43\textwidth}
    \centering
    \includegraphics[width=1.\columnwidth]{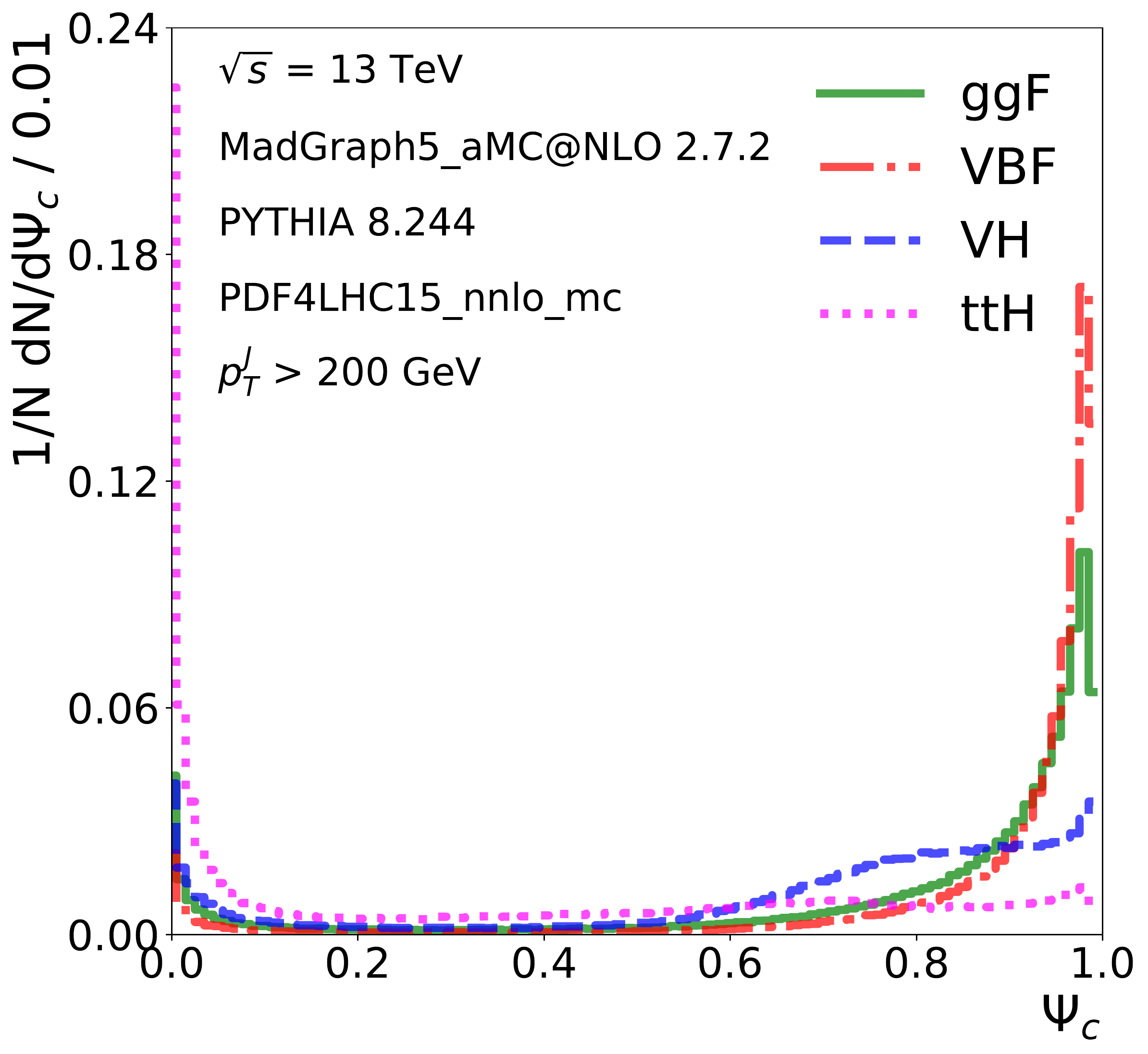}
    \end{subfigure}

\caption{Distributions of the five variables used in the BDT for each of the Higgs production modes. In all figures, $p^J_T$ represents the transverse momentum of the leading non-Higgs jet.}
\label{fig:Features}
\end{figure}

\noindent
The distributions of these five variables are shown in Fig. \ref{fig:Features}, in which the capability of each of the variables in discriminating between each production mode can be seen.  The salient features of these histograms are described below.

The distribution of leading jet invariant mass for the VH production mode has two peaks. The peak closest to 80 GeV is associated with the W boson jet and the other peak which is closest to 90 GeV is associated with the Z boson jet. For the ttH production, there is one bump and one peak. The peak is near 172 GeV because it comes from the top quark jet, while the bump comes from the W boson jet. The branching ratio of $t\rightarrow bW$ is almost 100\%~\citep{Tanabashi:2018oca}. Therefore, the invariant mass of the leading jet is close to the W boson mass if the final state of the bottom quark is not clustered into this leading jet. In the VBF and ggF production modes, the invariant mass of the leading jet in ggF events is generally higher than the mass of the leading jet in VBF events.

$|\Delta\eta|$ and $M_{JJ}$ distributions show a special property of VBF events: two forward jets. This is a powerful feature to distinguish VBF production from other production modes. 

The observable girth quantifies the momentum density inside the jet. A girth close to 0 means that the constituents of the jet concentrate around the axis of the jet. In contrast, if girth is close to 1, the constituents of jet are directed away from the jet's axis. Girth distributions for the VBF and ggF tend to be small, while those for the W/Z jet in the VH and the top jet in the ttH tend to be larger because the W/Z and top quark decay into more than one final particle with wider opening angle.

\subsection{The Two-stream Convolutional Neural Network}
\begin{figure}[t!]
\centering
\includegraphics[scale=0.20]{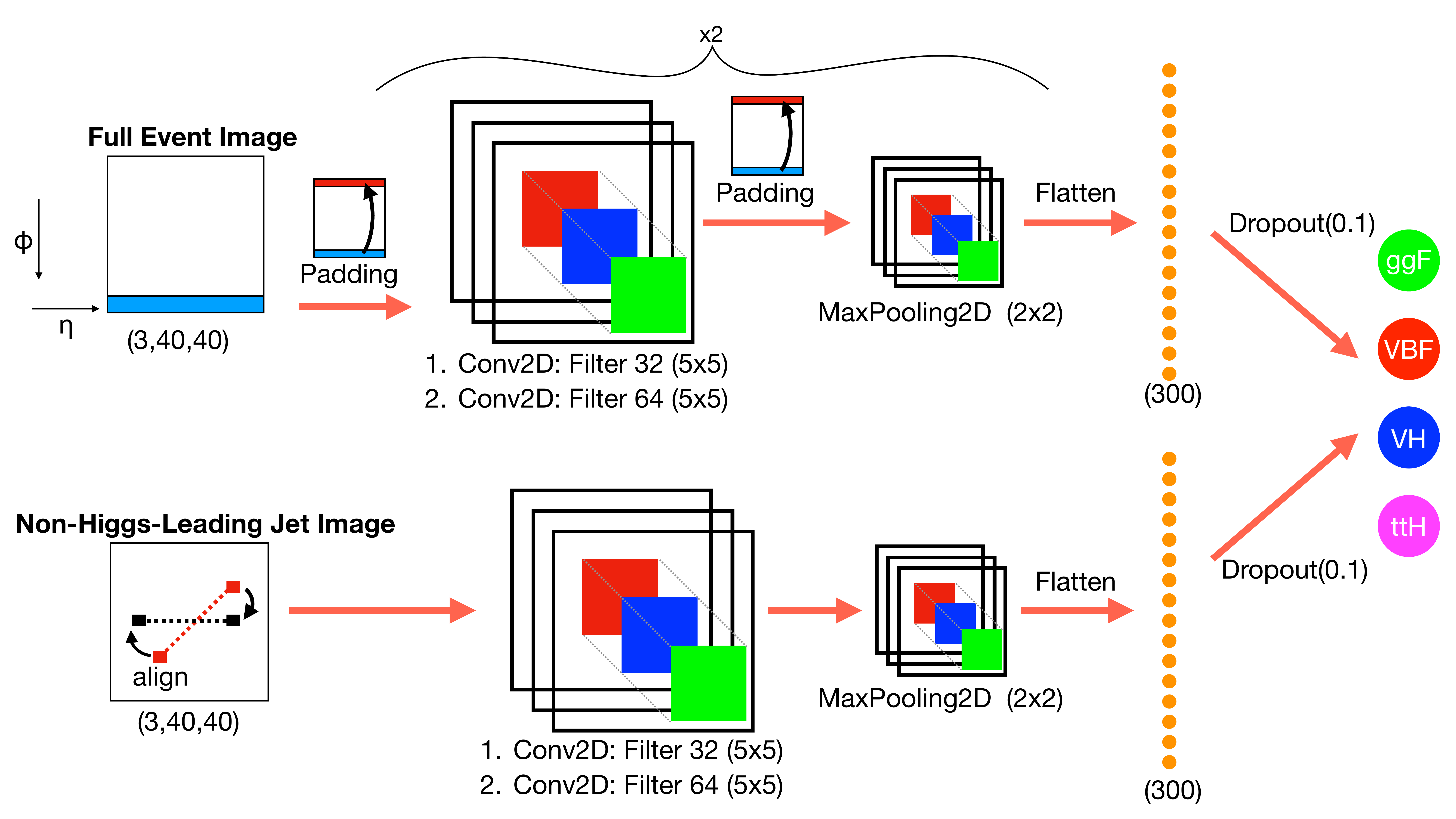}
\caption{Architecture of 2CNN, based on Ref.~\citep{Lin:2018cin}. The first stream (top) is used to process full-event images. The second stream (bottom) uses the information from the leading non-Higgs jet.}
\label{fig:Architecture of 2CNN}
\end{figure}

The 2CNN in this study is based on Ref.~\citep{Lin:2018cin}. One stream of the 2CNN is dedicated to global full-event information. The other stream is dedicated to processing local information in the leading non-Higgs jet. In addition, there are four outputs in the last layer for the four Higgs production modes. The two-stream architecture is shown schematically in Fig.~\ref{fig:Architecture of 2CNN}.

Details of the 2CNN are as follows. The convolution filter is 5$\times$5 in both streams, the maximum pooling layers are 2$\times$2, and the stride length is 1. Rectified linear unit (ReLU) activation functions are used for all intermediate layers of the neural network (NN). The first convolution layer in each stream has 32 filters and the second convolution layer in each stream has 64 filters. There are 300 neurons for the dense layer at the end of each stream. The two dense layers from each stream are fully connected to four output neurons with the softmax activation function $e^{x_i}/\sum_{i=1}^4 e^{x_i}$, which is the multidimensional generalization of the sigmoid.  The AdaDelta optimizer~\citep{DBLP:journals/corr/abs-1212-5701} is used to select the network weights. Between the last dense layer and output layer, Dropout~\citep{JMLR:v15:srivastava14a} regularization is added to reduce overfitting with the dropout rate = 0.1. The categorical cross entropy loss function is optimized in the NN training. For effectively utilizing the full information of the detector in the $\phi$ direction, a padding method is used to take the information in the bottom four rows of the input images and append them onto the top of the image. The {\texttt{Keras-2.3.0}} library is used to train a 2CNN model with the {\texttt{T{\footnotesize{ENSORFLOW}}-2.2.0}} \cite{tensorflow2015-whitepaper} backend, on a {\texttt{NVIDIA TITAN RTX 24 GB}}.

\begin{figure}[t!]
\centering
     \begin{subfigure}{1\textwidth}
        \centering
        \includegraphics[width=1\textwidth]{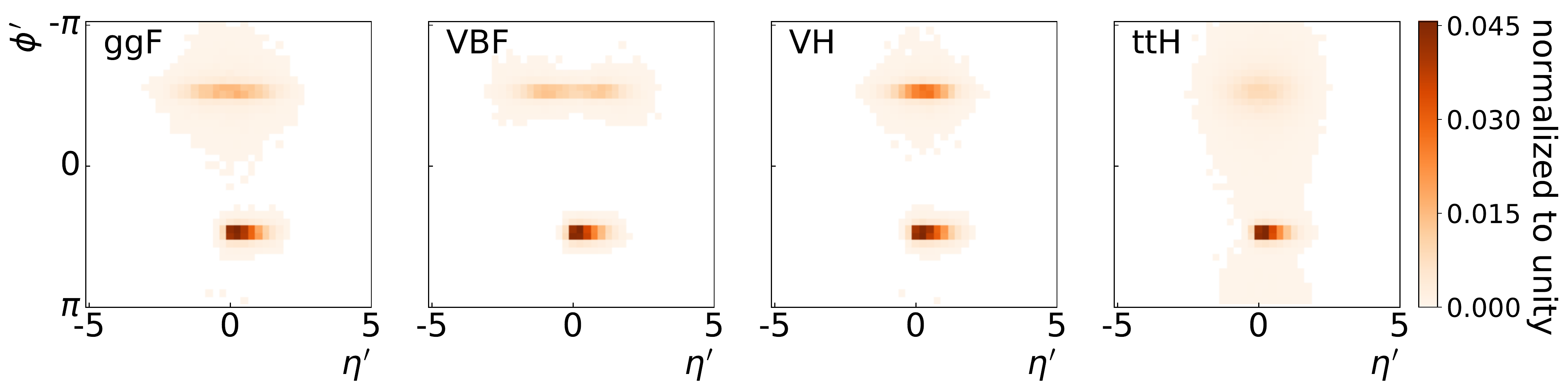}
     \end{subfigure}
     \begin{subfigure}{1\textwidth}
        \centering
        \includegraphics[width=1\textwidth]{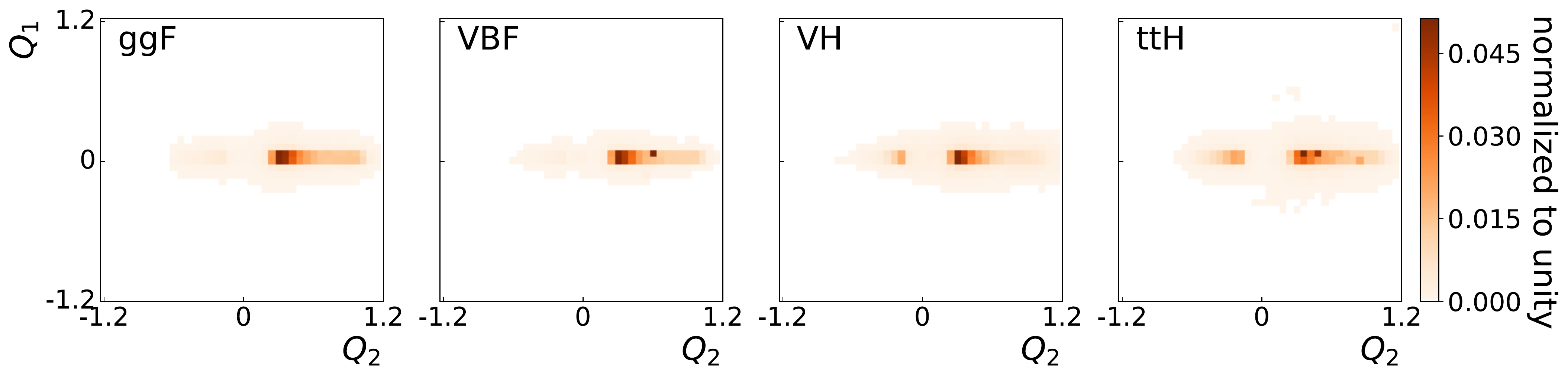}
     \end{subfigure}
\caption{The average of 10000 rotated full-event images (top) and leading non-Higgs jet images (bottom) in the charged $p_T$ channel. The coordinates $\phi'$ and $\eta'$ denote the new axis after the full-event images are rotated and flipped. $Q_1$ and $Q_2$ denote the new axes after the jet's axis is centralized and rotated. The intensity in each pixel is the sum of the charged particle $p_T$.  The total intensity in each image is normalized to unity. The resolution is 40$\times$40 pixels for each image.}
\label{fig:images_without_selection}
\end{figure}

The low-level inputs to the 2CNN are full-event images and the leading non-Higgs jet images. The resolution is 40$\times$40 pixels for both sets of images and jet images are in 1.5R$\times$1.5R range. The images consist of three channels, analogous to the Red-Green-Blue (RGB) channels of a color image. The pixel intensity for the three channels correspond to the sum of the charged particle $p_T$ , the sum of the neutral particle $p_T$ , and the number of charged particles in a given region of the image. There is no $p_T$ threshold for the contributions to pixel intensity. The full-event image covers effectively the entire $\eta$-$\phi$ cylinder ($|\eta|$ $<$ 5 ). 
Moreover, the full-event images are rotated so that the Higgs jet is always located at $\phi$ = $\pi$/2. Images are then flipped along the axis defined by $\eta$ = 0 to put the Higgs jet centroid in the region with positive $\eta$. The leading non-Higgs jet images are rotated so that the two subjets are aligned along the same axis. All images are normalized so that the intensities all sum to unity.  After normalization, the pixel intensities are standardized so that their distribution has a zero mean and unit variance.  Figure~\ref{fig:images_without_selection} shows the average full-event images and leading non-Higgs jet images in the charged $p_T$ channel. The patterns in the charged $p_T$ channel are similar to the other two channels.

\section{Results}\label{results}

The performance of the ML methods is quantified using receiver operating characteristic (ROC) curves.  A single classifier is formed for the ROC curves using p(ggF) and combining VBF, VH, and ttH together weighted by their relative cross sections.

\begin{figure}[t!]
\centering
	\begin{subfigure}{0.45\textwidth}
        \centering
        \includegraphics[width=2.5in]{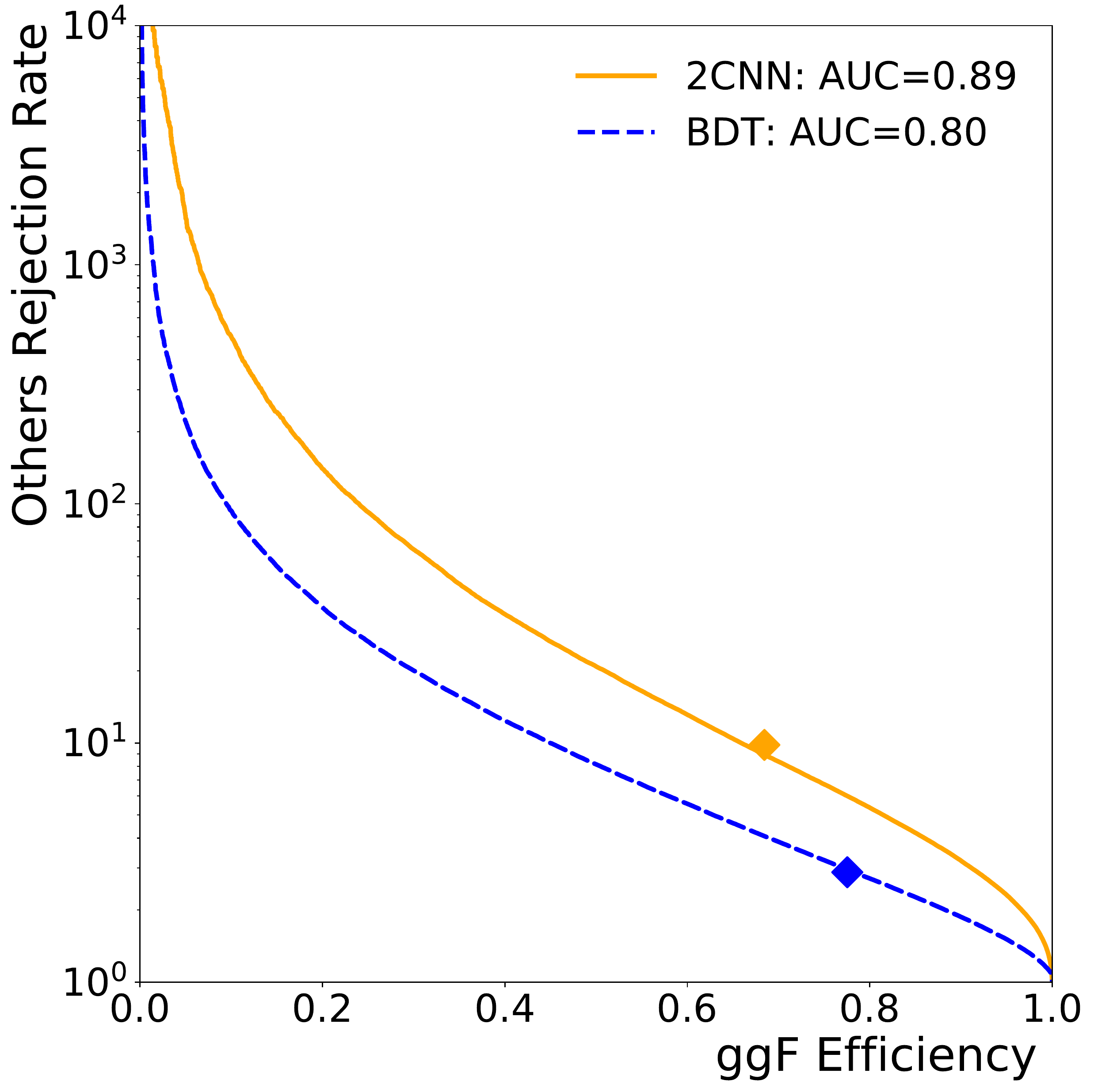}
     \end{subfigure}
     \begin{subfigure}{0.45\textwidth}
        \centering
        \includegraphics[width=2.5in]{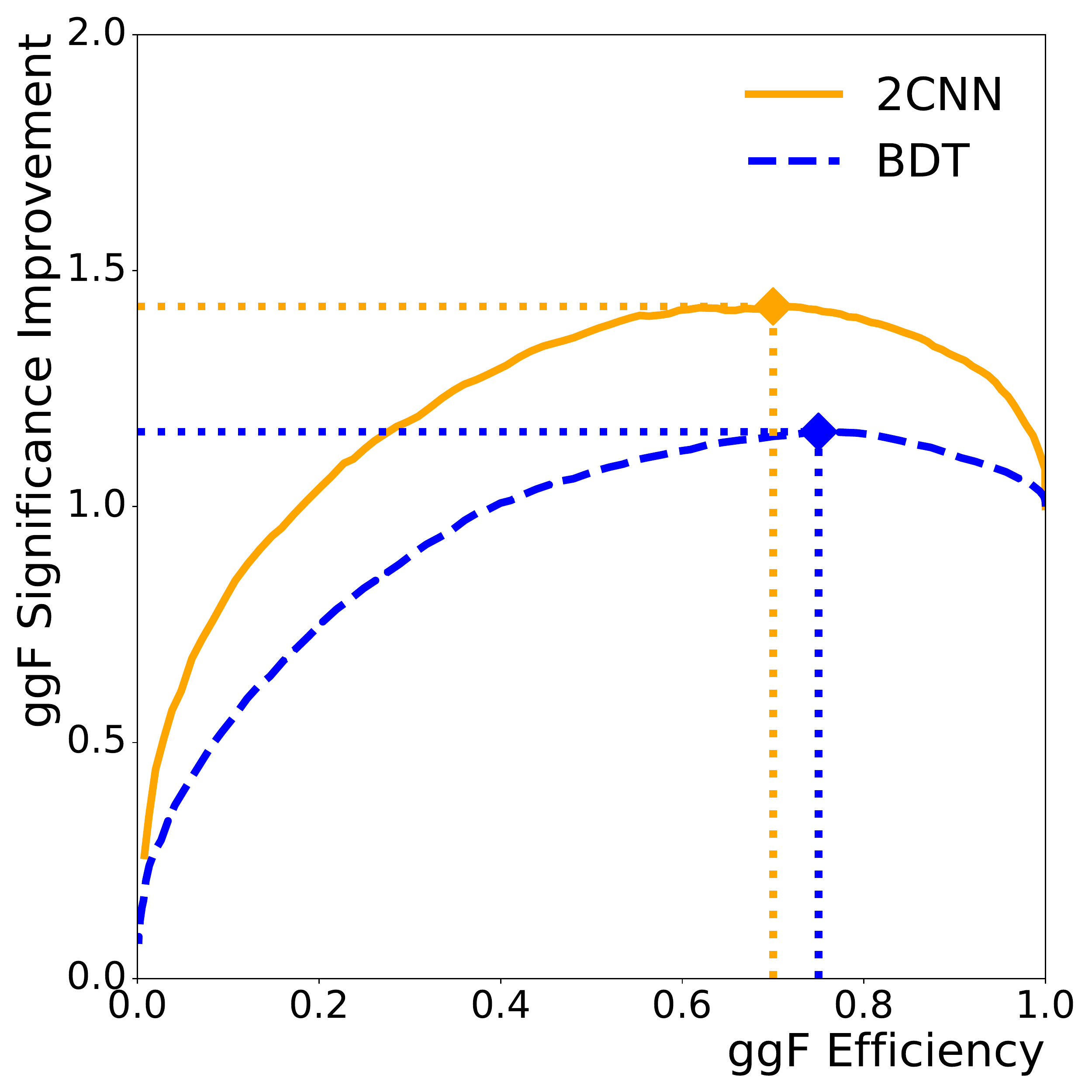}
     \end{subfigure}
\caption{Left: The ROCs for the BDT and the 2CNN, where each process is weighted by its relative cross section.  The diamonds are the working points at which the ggF significance improvement is maximized. Right: The ggF significance is normalized to its value at 100\% ggF efficiency. The right-hand side plot determines the working points.}
\label{fig:ROC_sig}
\end{figure}

The diamonds in the left plot of Fig. \ref{fig:ROC_sig} are working points of the BDT and the 2CNN where the ggF significance is maximized.  The ggF significance is calculated by $\sqrt{2[(s+b)ln(1+s/b)-s]}$, where $s$ is the number of the ggF event and $b$ is the number of event of the VBF+VH+ttH. At the working points, the ggF efficiency is 0.71 (0.78) and rejection rate of the other processes combined is 7.69 (2.86) for the 2CNN (BDT). For the significance improvement, the BDT has 1.2 and the 2CNN can reach up to 1.4. In the left plot of Fig. \ref{fig:ROC_sig}, the 2CNN shows better performance than the BDT. Therefore, it is interesting to look in more detail about the results and the performance of the 2CNN. The final results below will be shown at the 2CNN working point (the orange diamond).

\begin{figure}[t!]
\centering
     \begin{subfigure}{0.45\textwidth}
        \centering
        \includegraphics[width=2.5in]{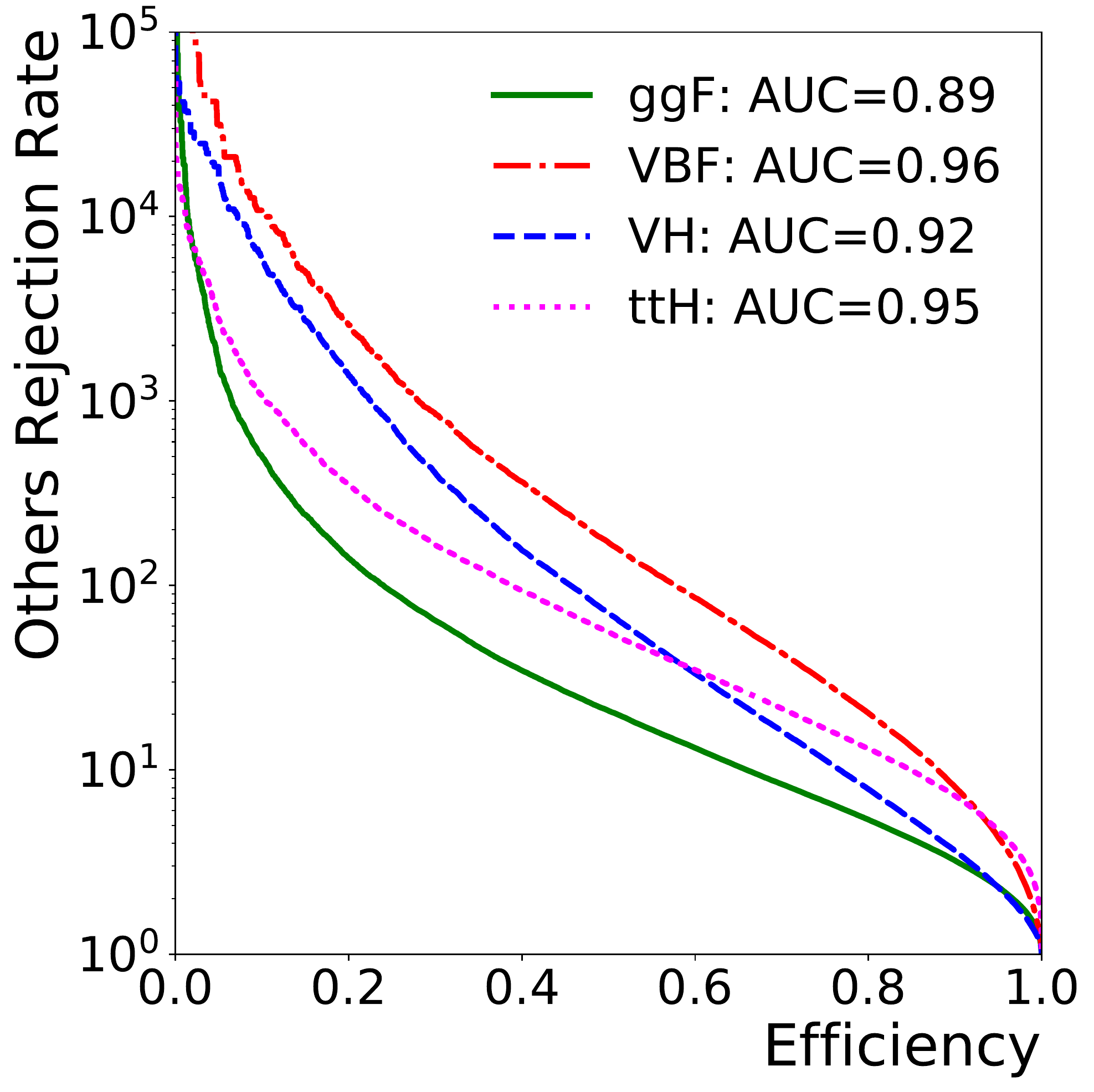}
     \end{subfigure}
     \begin{subfigure}{0.45\textwidth}
        \centering
        \includegraphics[width=2.5in]{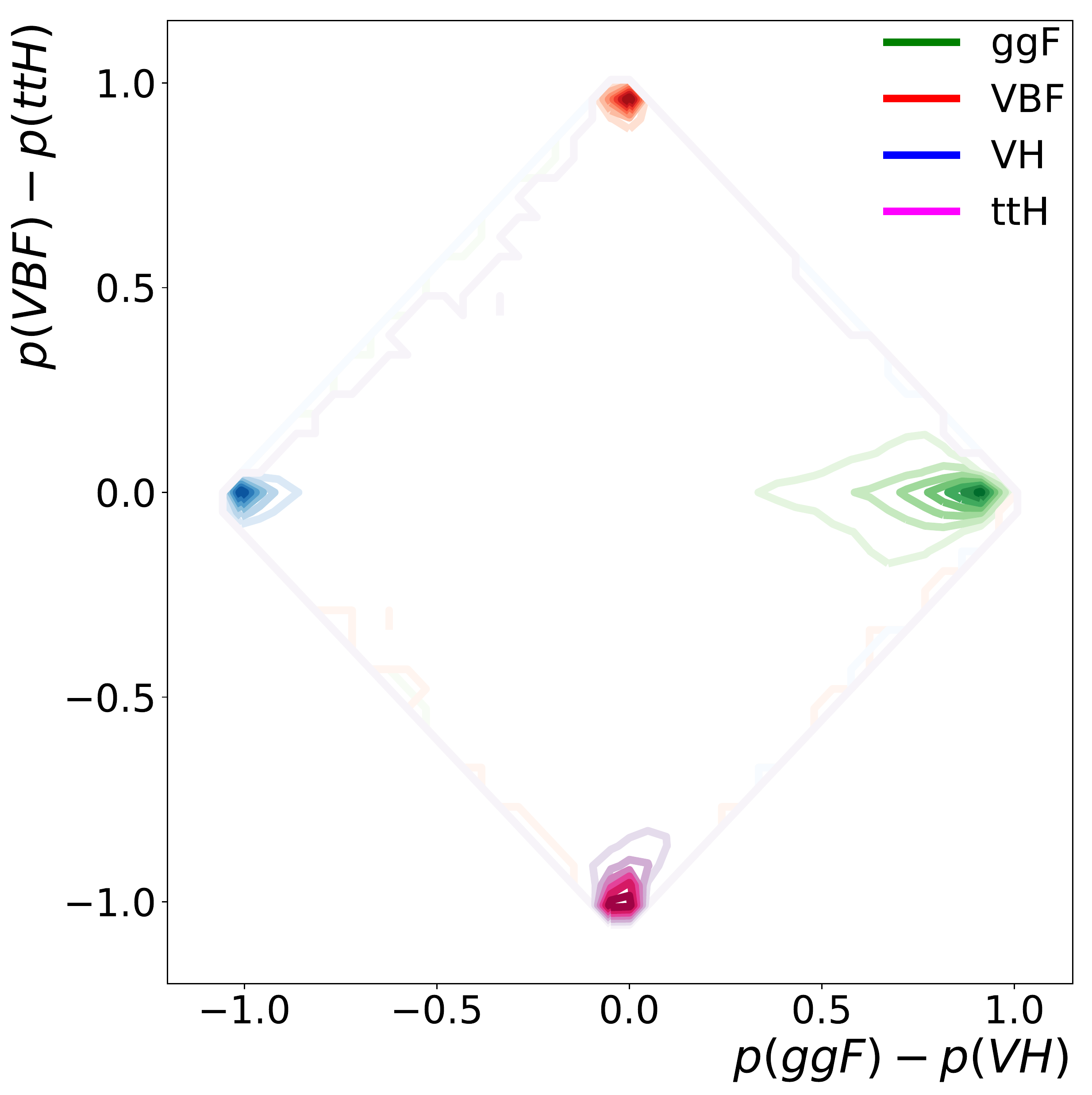}
     \end{subfigure}
     \begin{subfigure}{0.45\textwidth}
        \centering
        \includegraphics[width=2.5in]{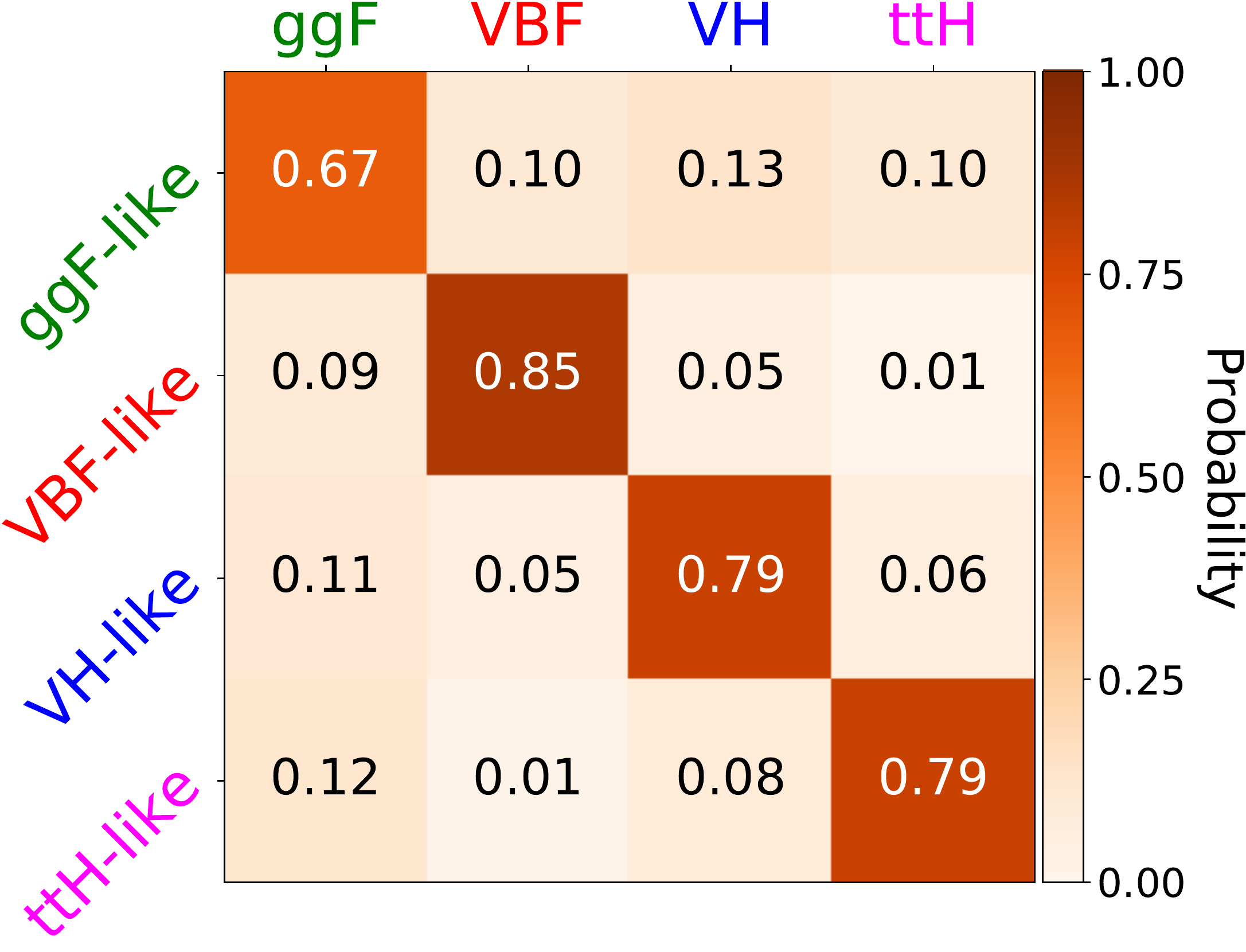}
     \end{subfigure}
     
\caption{Upper left: The x-axis is the efficiency of one of the four Higgs production modes, while the y-axis represents the others rejection rate which excludes the production mode in the x-axis. Upper right: A visualization of the differences in 2CNN output in 2-dimensional space. Each axis is the difference of two outputs. Bottom: Confusion matrix of the 2CNN. When the network randomly guesses, each component is 0.25.}
\label{fig:Performance_in_other_way}
\end{figure}

Figure~\ref{fig:Performance_in_other_way} illustrates three ways to quantify the 2NN performance. One is through use of ROC curves along different p(mode) axes. The second is the difference between p(VBF) and p(ttH) and the difference between p(ggF) and p(VH) in a 2-dimensional space, which is inspired by Ref.~\cite{Conway:2016caq,CMS-DP-2017-027}. The confusion matrix is a third performance metric.

The upper left plot of Fig.~\ref{fig:Performance_in_other_way} shows ROC curves along different signal axis, e.g. the green solid line is the weighted ROC when the ggF is signal and the other three processes are backgrounds. Along each p(mode) axis, the area under the ROC curve (AUC) is almost 0.90. 

The 2CNN model is trained to give values of (1,0,0,0) for ggF, (0,1,0,0) for VBF, (0,0,1,0) for VH, and (0,0,0,1) for ttH. From these four outputs, one can construct a 2-dimensional space that has the difference between the first and third outputs on the horizontal axis and the difference between the second and fourth outputs on the vertical axis. In the upper right plot of Fig.~\ref{fig:Performance_in_other_way}, the four Higgs production modes will distribute around (-1,0) for VH, (0,-1) for ttH, (0,1) for VBF and (1,0) for ggF.

The confusion matrix can also characterize the accuracy of classification. It shows how many true ggF, VBF , VH, and ttH are classified as such. From Fig.~\ref{fig:Performance_in_other_way}, bottom, the ggF mode is classified 67\% correctly and the other three Higgs production modes can reach 80\%. 

\begin{figure}[t!]
\centering
	\begin{subfigure}{0.45\textwidth}
        \centering
        \includegraphics[width=2.5in]{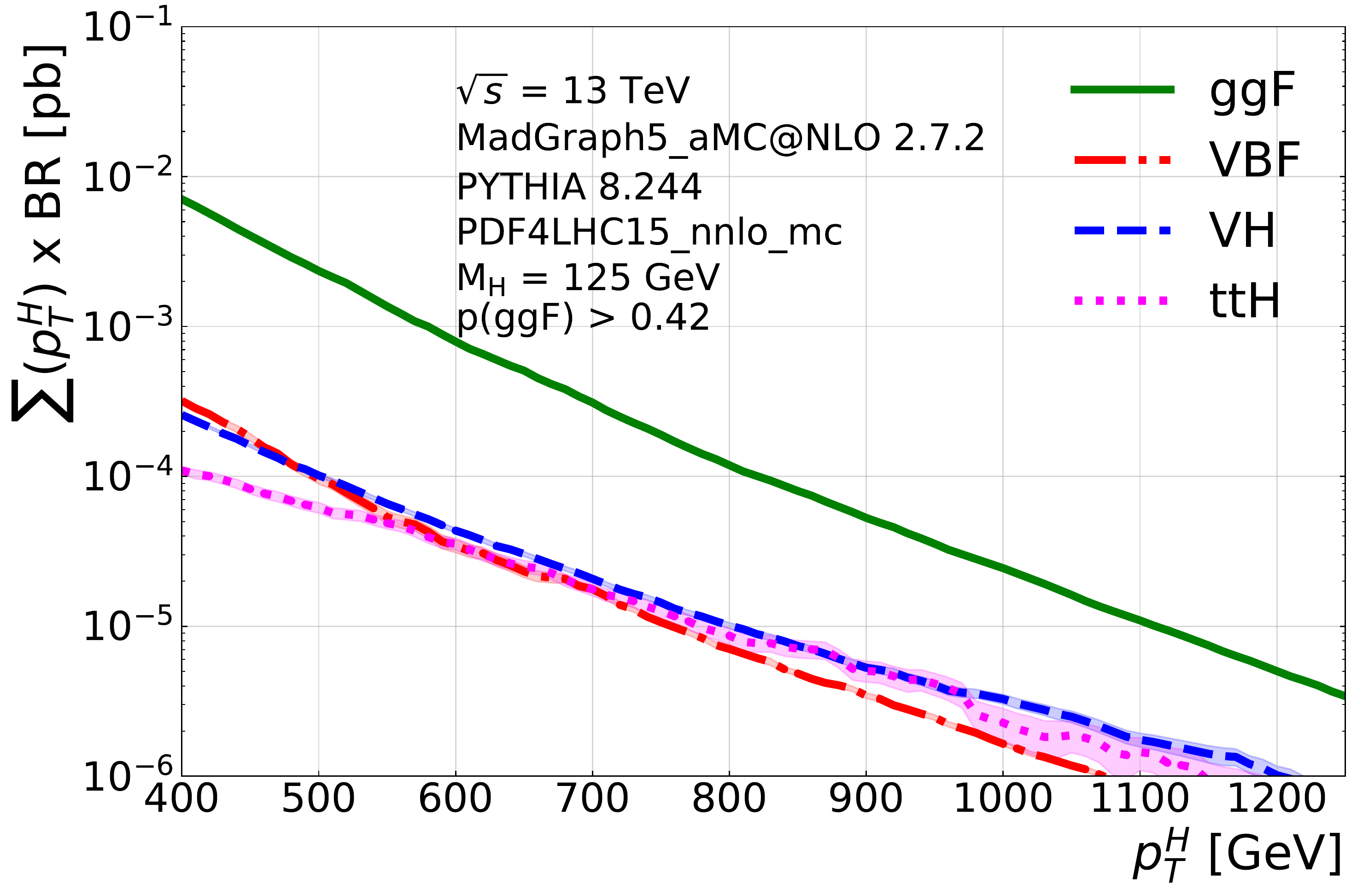}
        \caption{The cumulative cross section at working point.}
     \end{subfigure}
	\begin{subfigure}{0.45\textwidth}
        \centering
        \includegraphics[width=2.5in]{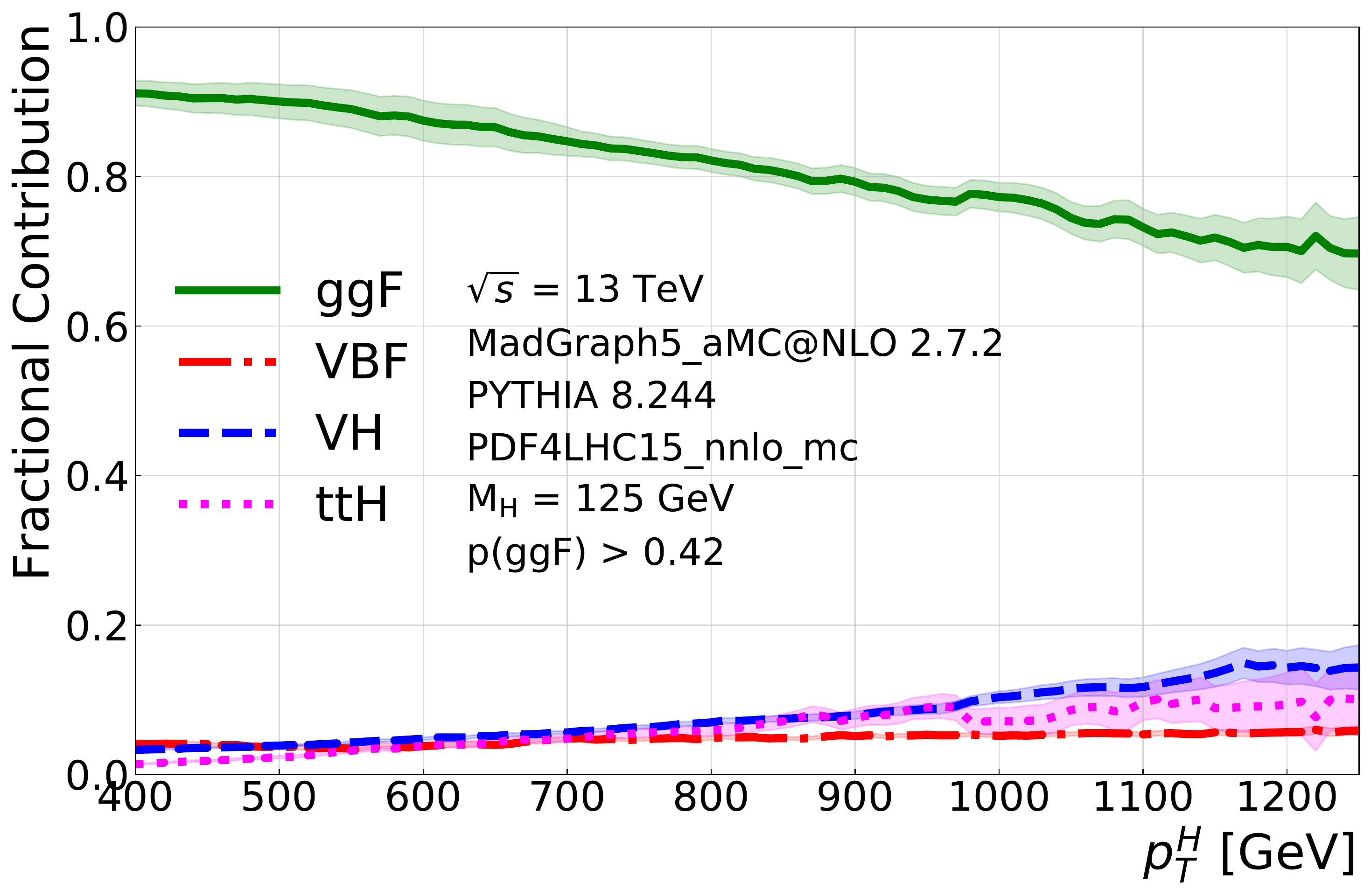}
        \caption{The fractional contribution at working point.}
     \end{subfigure}
\caption{The 2CNN performance in the cumulative cross section and fractional contribution. At the 2CNN working point, the ggF fraction can be highly increased across the whole boosted range. The lighter error band of each production in the figures is the one $\sigma$ statistical uncertainty.}
\label{fig:CumulativeXection_compare}
\end{figure}

Fig.~\ref{fig:CumulativeXection_compare} focuses on the ggF mode and presents the cumulative cross section and fractional contribution at the 2CNN working point. In the range 400 GeV $<$ $p^H_T$ $<$ 600 GeV and the highly boosted region (1000 GeV $<$ $p^H_T$ $<$ 1250 GeV), the fractional contribution of ggF can improve from 0.55 to 0.85 and from 0.4 to 0.6, respectively.

 Not only can the ggF production mode be clearly separated, but the 2CNN classifier also has the potential to improve the precision for other Higgs production modes in extreme regions of phase space. The results of other Higgs production modes are in Appendix \ref{Appendix}, which are shown in the cumulative cross section.

\section{Discussion} \label{discussion}
The 2CNN has been demonstrated to separate the four Higgs production modes effectively via the global full-event images and the local leading non-Higgs jet images. It is useful to look into what information the 2CNN relies on for classification. Following Ref.~\cite{Lin:2018cin,deOliveira:2015xxd}, several visualization methods are used for this purpose. The behavior is similar in all three image channels, so this section focuses on the charged $p_T$ only.

\begin{figure}[t!]
\centering
	\begin{subfigure}{1\textwidth}
        \centering
        \includegraphics[width=1\textwidth]{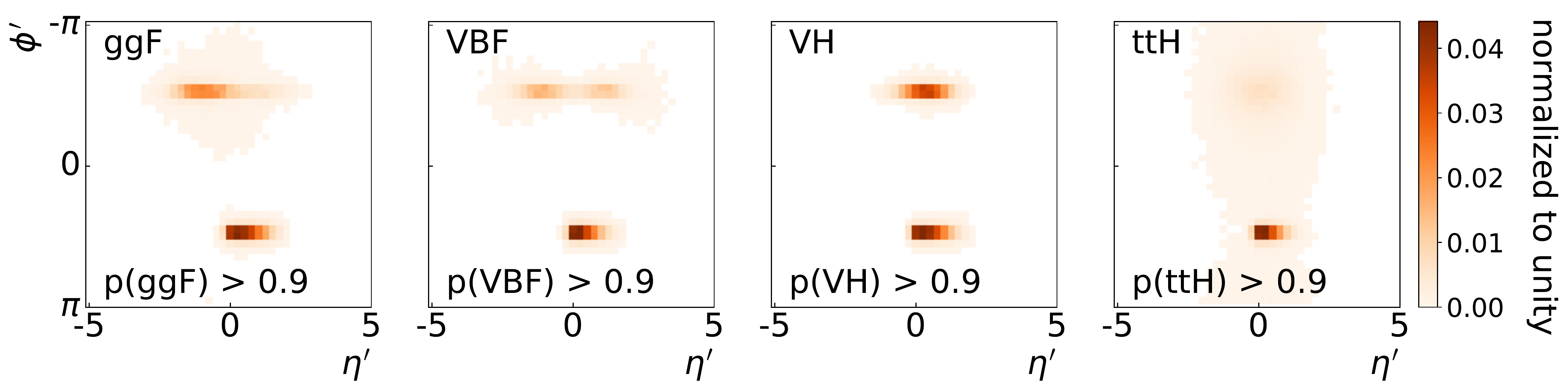}
     \end{subfigure}
\caption{Average rotated full-event images with high score in the charged particle $p_T$ channel. The intensity in each pixel is the sum of the charged particle $p_T$. Total intensity in each image is normalized to unity. }
\label{fig:full_event_image_charge_pt}
\end{figure}

The full-event images are rotated to remove the $\phi$ symmetry inherent to the LHC. More physics information can be easily recognized in the full-event images. Fig.~\ref{fig:full_event_image_charge_pt} illustrates the global full-event images of the four Higgs production modes with high-score (the 2CNN output~$>$~0.9). The leading non-Higgs jet's substructures are clear in the upper region of each sub-figure. The structures are from the gluon jet in the ggF, two-forward-quark jets in the VBF, vector boson jet in the VH, and two top jets in the ttH. The Higgs jets are in the lower regions and contain the same patterns in the four modes. Therefore, the 2CNN can extract that global information from the rotated full-event images. 

\begin{figure}[t!]
\centering
	\begin{subfigure}{1\textwidth}
        \centering
        \includegraphics[width=0.9\textwidth]{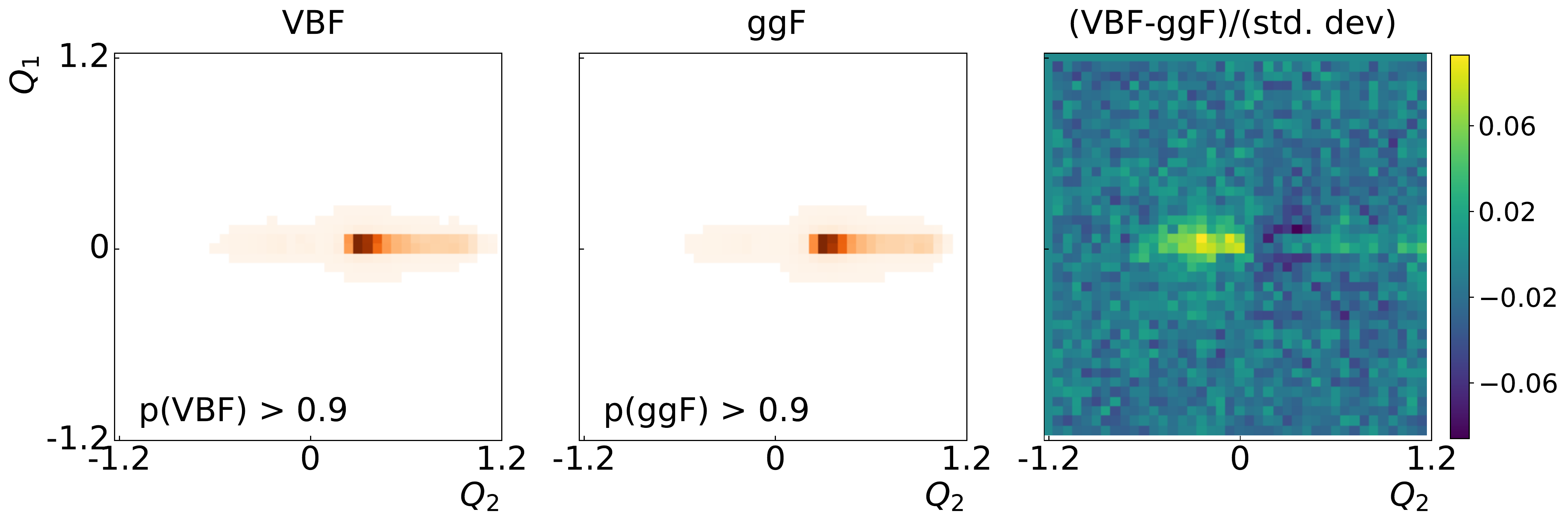}
     \end{subfigure}
     \hspace{0.9\textwidth}
     \begin{subfigure}{1\textwidth}
        \centering
        \includegraphics[width=0.9\textwidth]{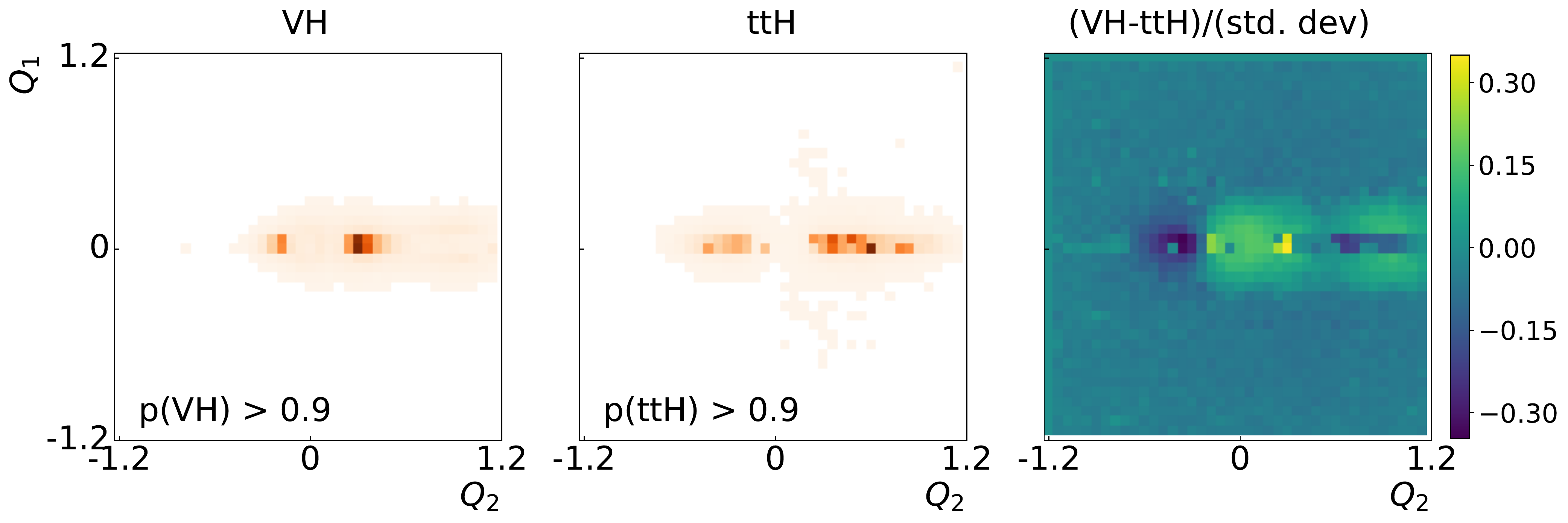}
     \end{subfigure}
\caption{Average leading non-Higgs jet images with a high score in the charged particle $p_T$ channel (left column and middle column). The average difference between the VBF and ggF (upper right) and the average difference between the VH and ttH (bottom right) leading non-Higgs jet images. $Q_1$ and $Q_2$ denote the new axes after the jet's axis is centralized and rotated.}
\label{fig:jet_image_difference}
\end{figure}

\begin{figure}[h]
\centering
	\begin{subfigure}{0.4\textwidth}
        \centering
        \includegraphics[width=1\textwidth]{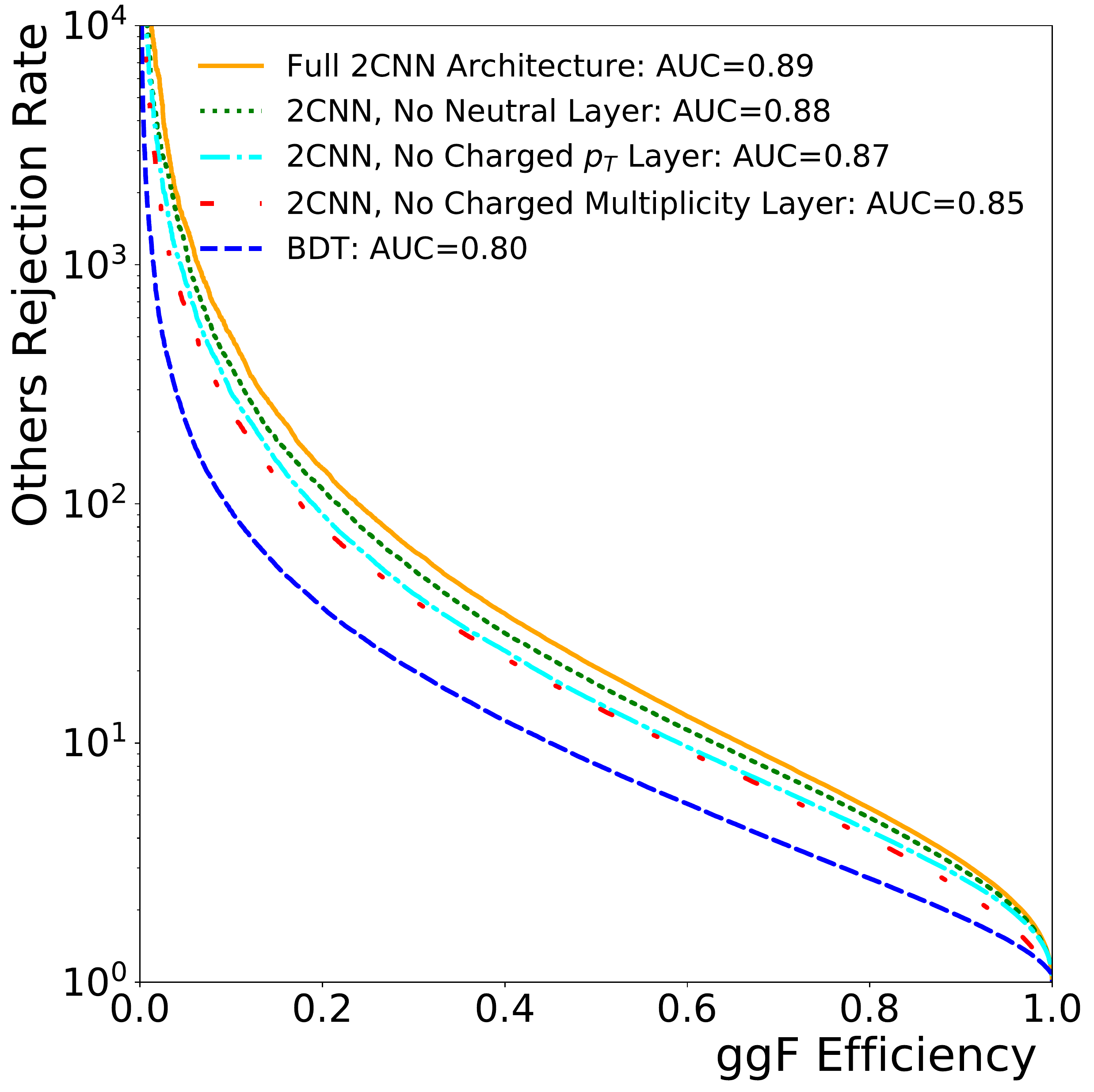}
    \end{subfigure}
    \begin{subfigure}{0.4\textwidth}
        \centering
        \includegraphics[width=1\textwidth]{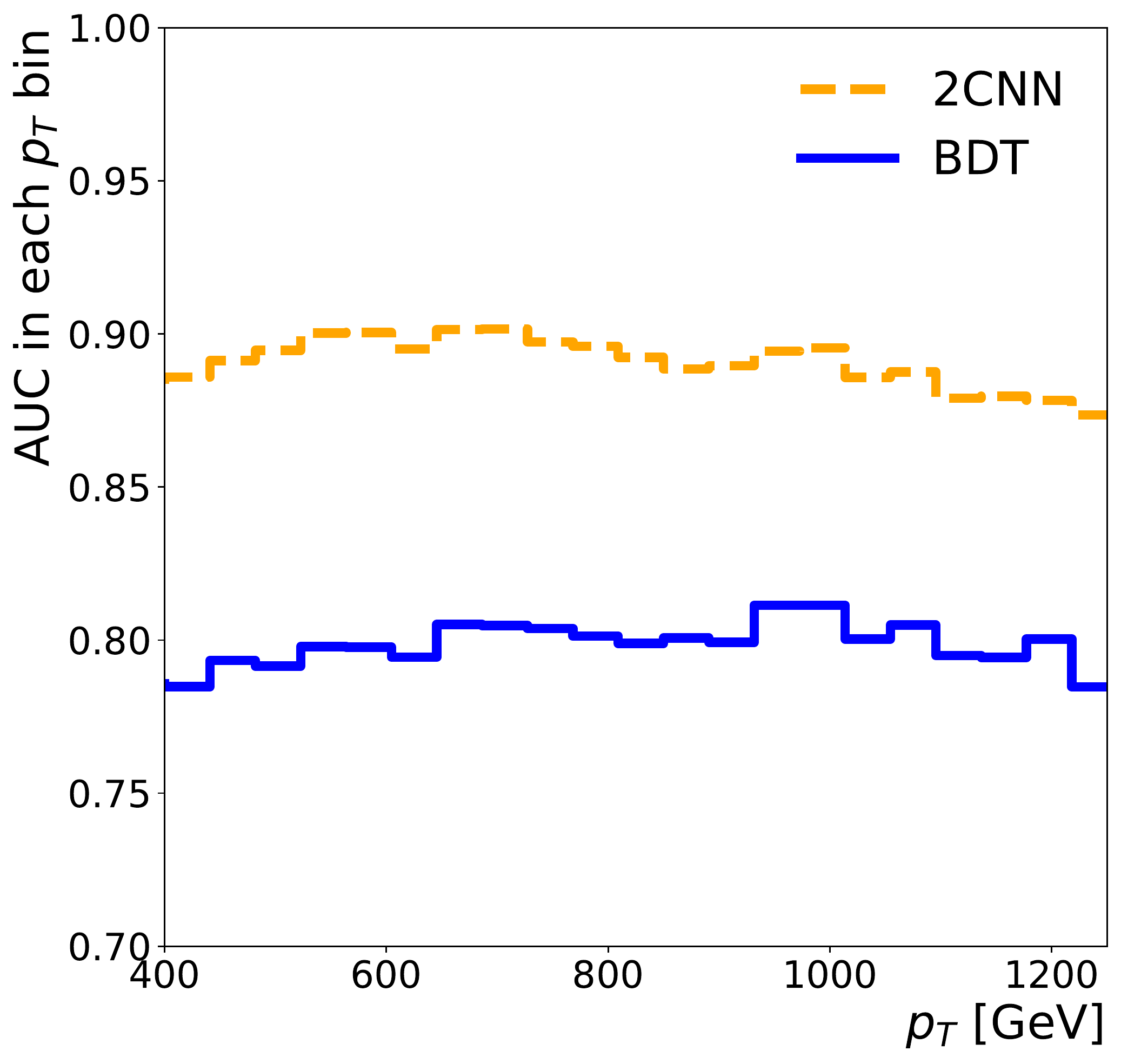}
     \end{subfigure}
\caption{Left: The performance for various neural network setups. Right: The AUC performance in each $p_T$ bin with 40 GeV bin size.}
\label{fig:various_setups}
\end{figure}

Furthermore, Fig.~\ref{fig:jet_image_difference} shows the average leading non-Higgs jet images with a high score (the 2CNN's output~$>$~0.9) for each mode and the difference between modes. The 2CNN seems to be using the same strategy to identify the leading non-Higgs jet for the ggF and VBF. The average leading non-Higgs jet images with a high score for the VBF and ggF look very similar, but the difference between them still reveals interesting structure. This additional information can be observed in the upper right plot of Fig.~\ref{fig:jet_image_difference}. In particular, it can be seen in the left half of the top right panel that the leading non-Higgs jet in the VBF mode tends to be harder than the ggF due to in part to the flavor composition of the jets. However, the ggF mode tends to deposit slightly more energy than the VBF mode around the core area of jets.
For the VH and ttH, although they also contain two-prong structure, the non-Higgs-jet images of the VH show more $p_T$ in wider prongs than those of the ttH. The top quark in the ttH decays into a W boson and a b quark, so the two-prong structure tends to have more separated than in the VH case.

The left panel of Fig.\ref{fig:various_setups} shows the the performance is relatively unchanged when the neutral layer is removed. The charged $p_T$ and charged multiplicity layer are rather insensitive to pileup effects. Therefore, the 2CNN will be relatively stable to various experimental effects such as pileup. to various

The right panel of Fig.\ref{fig:various_setups} shows that in the $p_T$ range between 400 to 1250 GeV with 40 GeV bin size, BDT and 2CNN classifiers have uniform performance based on the AUC metric. It implies that these two classifiers perform uniform disentangling power in each $p_T$ bin.


\section{Conclusions and Outlook}\label{conclusions}

BDT and 2CNN techniques have been studied for disentangling the four main boosted Higgs production modes with $\mathrm{b\bar{b}}$ decay at the LHC. The goal was to identify ggF production accurately, given its sensitivity in the study of BSM from the $gg\to \mathrm{H}$ loops at high $p_T$, and it has been shown that the 2CNN method can significantly enhance the separation between ggF and other modes.

The 2CNN architecture in this study is built on the proposal from Ref.~\cite{Lin:2018cin}. To generalize the 2CNN to the case of four Higgs production mode classification, the architecture used in this work has 4-class outputs and contains one stream acting on global event information, and the other stream acting on information from the leading non-Higgs jet. This approach is amenable to visualizations that can provide some insight into what the NN is using for classification. 

While the focus in this paper has been on enhancing ggF, the 2CNN approach can also be used to enhance other production modes as well. For example, the 2CNN technique can make the VBF and VH fractional contributions reach 77\% and 78\% with $p^H_T$ threshold = 400 GeV, respectively. This could be useful for precision measurements of electroweak symmetry breaking in the boosted region.  Another possible application is for the Heavy Vector Triplet (HVT) model, which predicts a Heavy Vector that can decay into the VH mode \cite{Pappadopulo:2014qza}. This kind of VH production should produce boosted Higgs, meaning the clear separation of the VH in the boosted case via ML techniques can aid the search for the HVT. In addition, the top quark Yukawa coupling in the Higgs precision measurement also can be studied in the high-$p_T$ Higgs process through extracting the ttH from other production modes~\citep{Azatov:2013xha}. 

In summary, we have applied a deep CNN to incorporate both local and global information for boosted Higgs boson identification. Additionally, we have studied a boosted decision tree, which is also effective, but is less powerful than the neural network.  Previous work showed that a similar neural network architecture could significantly enhance the presence of the Higgs signal over generic multijet backgrounds.  We have shown that this technique has a great potential to further enhance the discovery of boosted Higgs bosons via ggF (or other topologies) by precisely separating events into various production modes.  This approach is flexible and may be able to enhance our sensitivity to BSM in a variety of final states with and without Higgs bosons at the LHC and beyond.

\newpage
\appendix
\section{Appendix}\label{Appendix}

\begin{figure}[h]
\centering
	\begin{subfigure}{0.45\textwidth}
        \centering
        \includegraphics[width=2.5in]{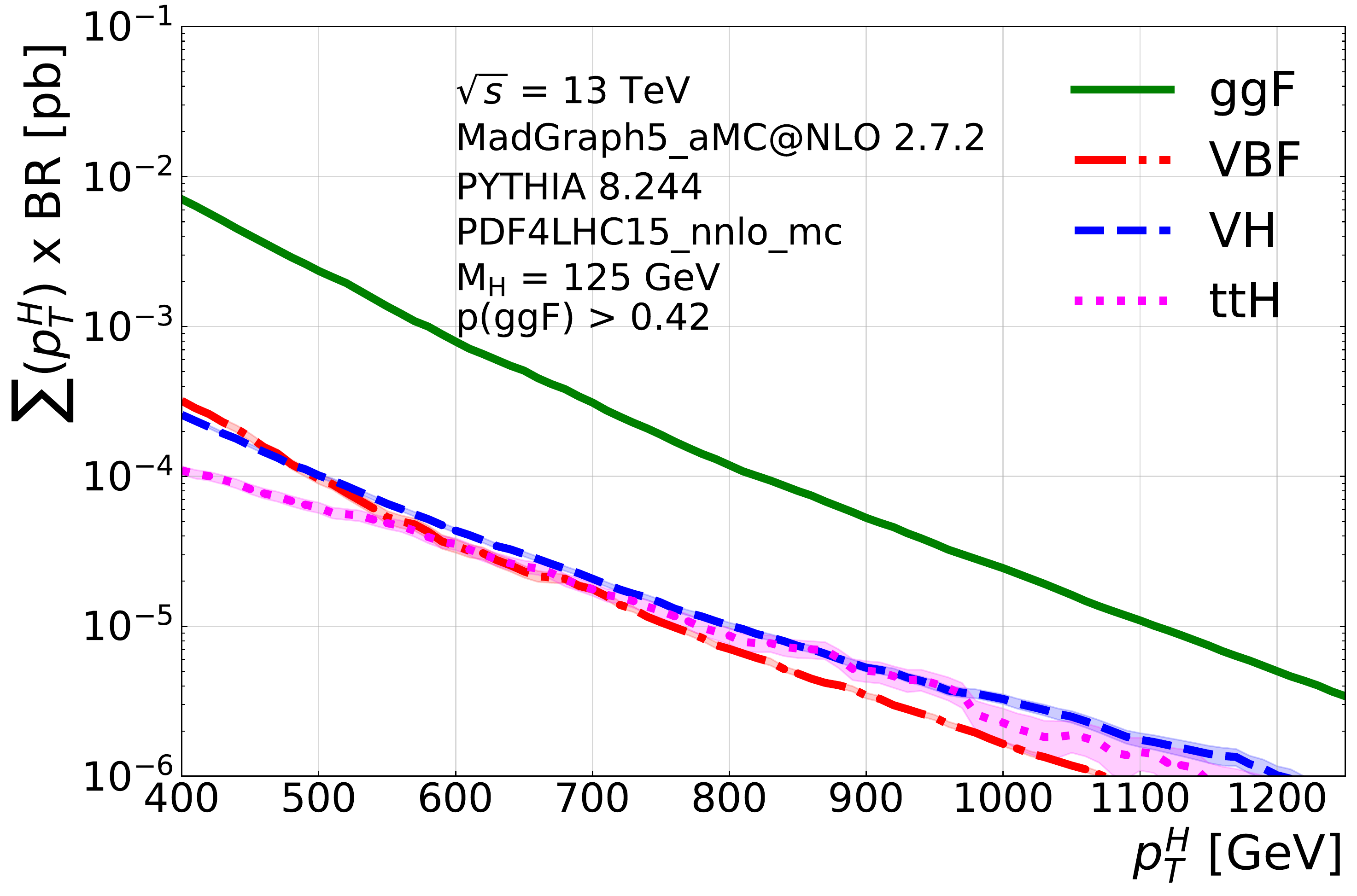}
        \caption{The cumulative cross section at ggF working point.}
     \end{subfigure}
     \begin{subfigure}{0.45\textwidth}
        \centering
        \includegraphics[width=2.5in]{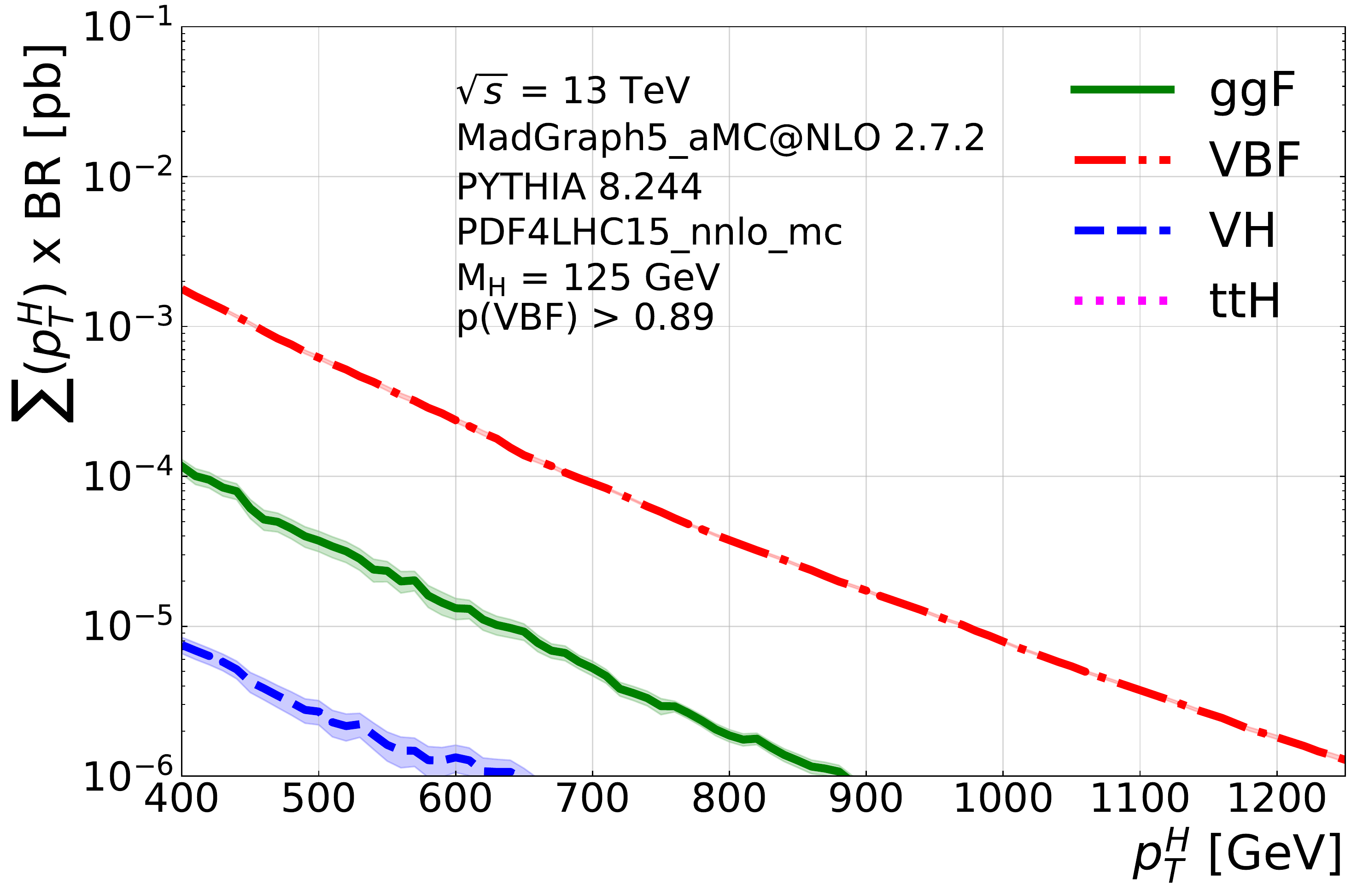}
        \caption{The cumulative cross section at VBF working point.}
     \end{subfigure}
     \begin{subfigure}{0.45\textwidth}
        \centering
        \includegraphics[width=2.5in]{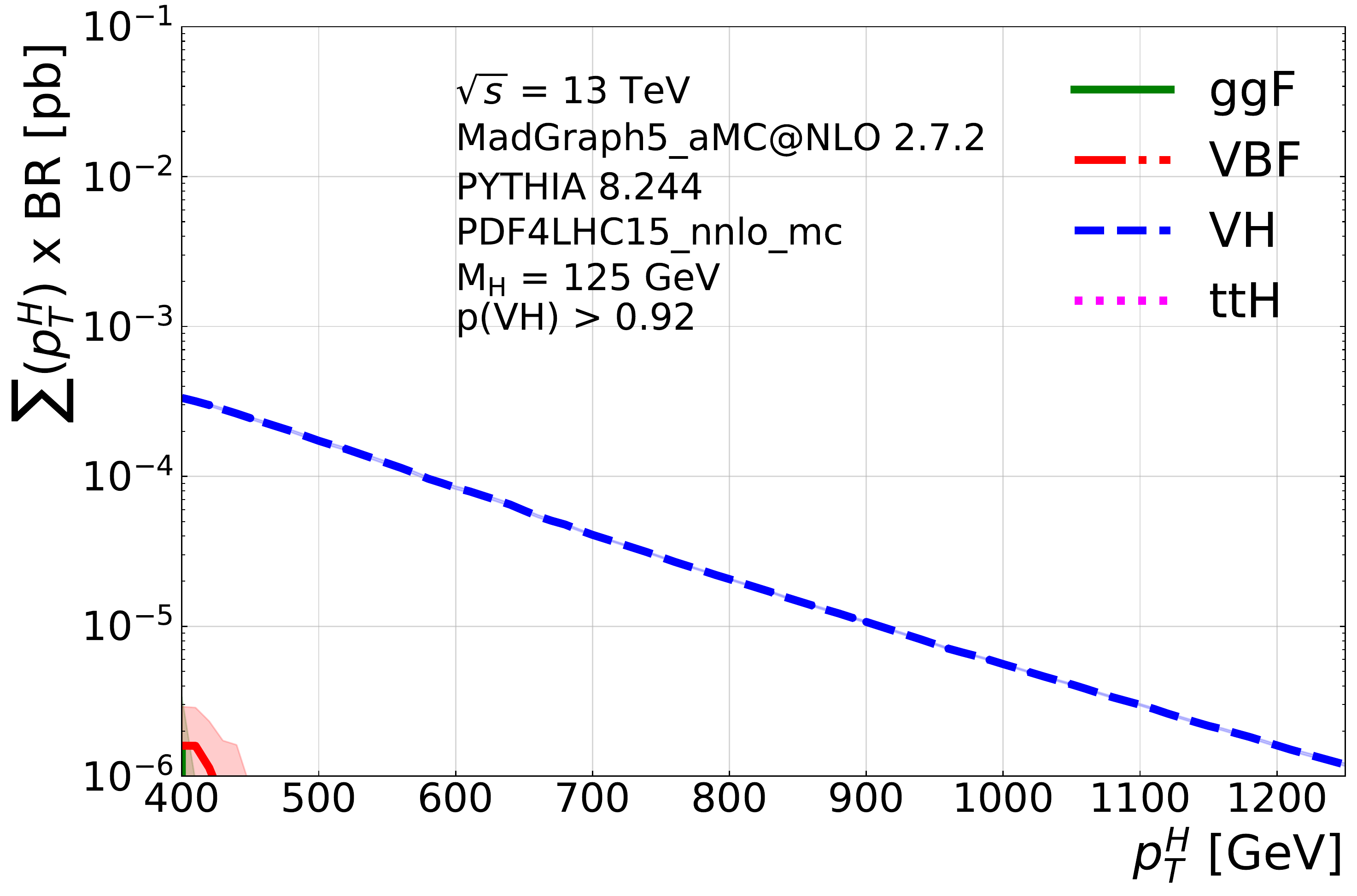}
        \caption{The cumulative cross section at VH working point.}
     \end{subfigure}
     \begin{subfigure}{0.45\textwidth}
        \centering
        \includegraphics[width=2.5in]{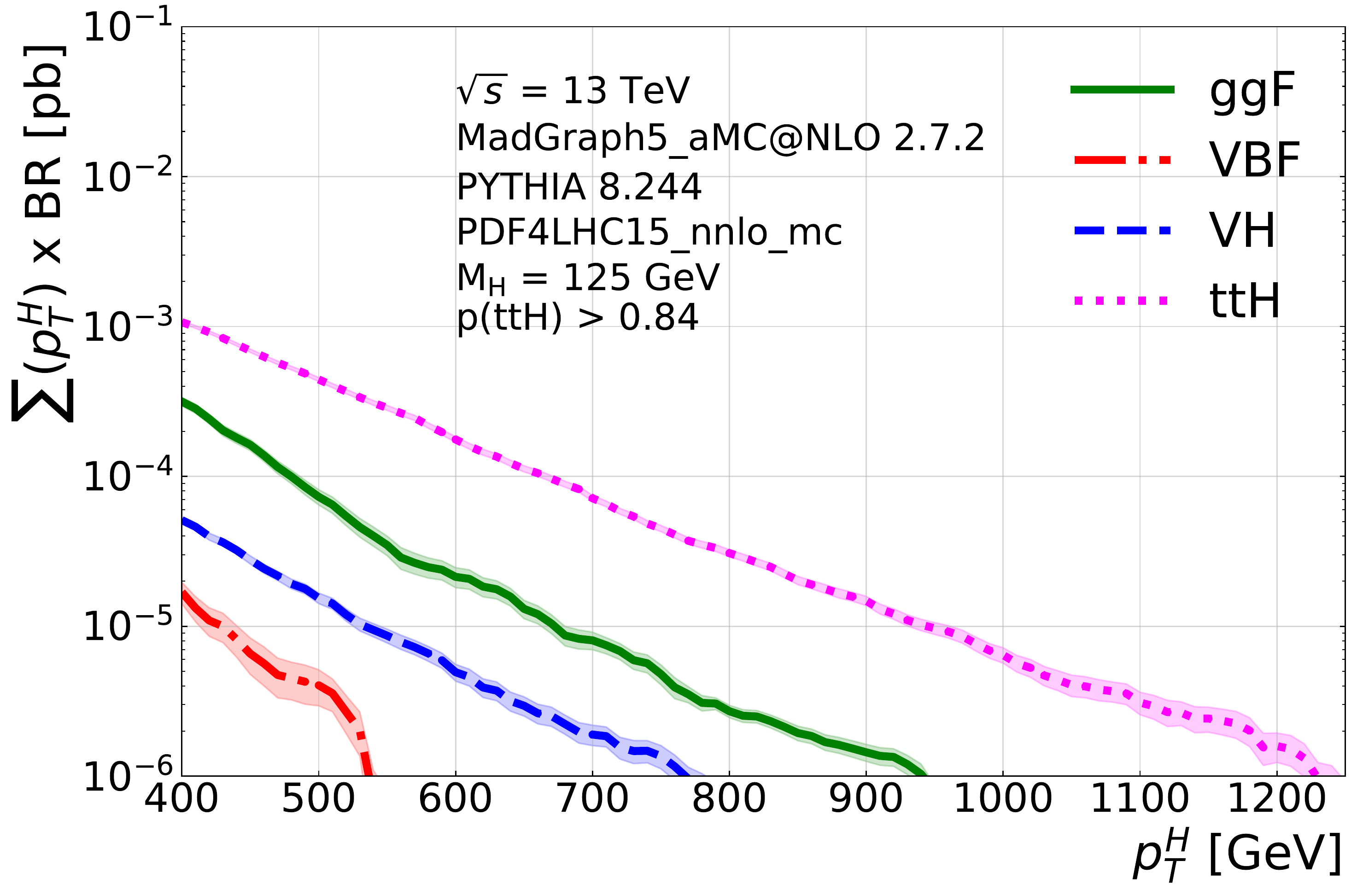}
        \caption{The cumulative cross section at ttH working point.}
     \end{subfigure}
\caption{The 2CNN performance in the cumulative cross section at working point of each production mode.}
\label{fig:performance_for_each_axis}
\end{figure}

Fig. \ref{fig:performance_for_each_axis} shows disentangling performance along p(ggF), p(VBF), p(VH) and p(ttH). In each Higgs production axis, the 2CNN architecture can provide exceptionally clear separation in boosted region. 

\newpage
\acknowledgments

We thank Lanke Fu for collaboration during early stages of this work. 
We also thank Carter Vu and John Spencer for their detailed suggestions on the text. Furthermore, we are grateful to Joshua Lin for help with the machine learning setup as well as detailed feedback on the manuscript.  Additionally, we thank Frank Tackmann for useful discussions and the organizers of Les Houches 2019 where the idea for this project originated. We also thank Ian Moult and Kingman Cheung for useful discussions and encouragement.  This work was supported by the U.S. Department of Energy, Office of Science under contract DE-AC02-05CH11231. S.-C. Hsu is supported by the U.S. Department of Energy, Office of Science, Office of Early Career Research Program under Award number DE-SC0015971. The work of Y.-L.C. was supported by the Taiwan MoST with the grant number MOST-107-2112- M-007-029-MY3.

\newpage
\bibliographystyle{jhep}
\bibliography{references,HEPML}

\end{document}